\documentclass[aps,prx,twocolumn,reprint,superscriptaddress]{revtex4-1}

\usepackage{times,ulem}
\usepackage{amsmath,amssymb}
\usepackage{bm}
\usepackage{graphicx}
\usepackage{color}
\usepackage[dvipsnames]{xcolor}
\usepackage{hyperref,array,dsfont}
\usepackage{qcircuit,braket}

\begin{document}

\title{Split-ring polariton condensates as  macroscopic two-level quantum systems}

\author{Yan Xue}
\email{xy4610@jlu.edu.cn}
\affiliation{Westlake University, School of Science, 18 Shilongshan Road, Hangzhou 310024, Zhejiang Province, China}
\affiliation{College of Physics, Jilin University, Changchun 130012, China}
\affiliation{Westlake Institute for Advanced Study, Institute of Natural Sciences, 18 Shilongshan Road, Hangzhou 310024, Zhejiang Province, China}

\author{Igor Chestnov}
\email{igor\_chestnov@westlake.edu.cn}
\affiliation{Westlake University, School of Science, 18 Shilongshan Road, Hangzhou 310024, Zhejiang Province, China}
\affiliation{Westlake Institute for Advanced Study, Institute of Natural Sciences, 18 Shilongshan Road, Hangzhou 310024, Zhejiang Province, China}
\affiliation{Vladimir State University, Gorkii St. 87, 600000, Vladimir, Russia}

\author{Evgeny Sedov}
\affiliation{Westlake University, School of Science, 18 Shilongshan Road, Hangzhou 310024, Zhejiang Province, China}
\affiliation{Westlake Institute for Advanced Study, Institute of Natural Sciences, 18 Shilongshan Road, Hangzhou 310024, Zhejiang Province, China}
\affiliation{Vladimir State University, Gorkii St. 87, 600000, Vladimir, Russia}

\author{Evgeny Kiktenko}
\affiliation{Russian Quantum Center, Skolkovo, Moscow 143025, Russia}
\affiliation{Moscow Institute of Physics and Technology, Dolgoprudny 141700, Russia}

\author{Aleksey Fedorov}
\affiliation{Russian Quantum Center, Skolkovo, Moscow 143025, Russia}
\affiliation{Moscow Institute of Physics and Technology, Dolgoprudny 141700, Russia}

\author{Stefan Schumacher}
\affiliation{Department of Physics and Center for Optoelectronics and Photonics Paderborn (CeOPP), Universit\"{a}t Paderborn, Warburger Strasse 100, 33098 Paderborn, Germany}

\author{Xuekai Ma}
\email{xuekai.ma@gmail.com}
\affiliation{Department of Physics and Center for Optoelectronics and Photonics Paderborn (CeOPP), Universit\"{a}t Paderborn, Warburger Strasse 100, 33098 Paderborn, Germany}

\author{Alexey Kavokin}
\email{a.kavokin@westlake.edu.cn}
\affiliation{Westlake University, School of Science, 18 Shilongshan Road, Hangzhou 310024, Zhejiang Province, China}
\affiliation{Westlake Institute for Advanced Study, Institute of Natural Sciences, 18 Shilongshan Road, Hangzhou 310024, Zhejiang Province, China}

\begin{abstract}
Superposition states of circular currents of exciton-polaritons mimic the superconducting flux qubits. The phase of a polariton fluid must change by an integer number of ${2\pi}$, when going around the ring. If one introduces a ${\pi}$-phase delay line in the ring, the fluid is obliged to propagate a clockwise or anticlockwise circular current to reduce the total phase gained over one round-trip to zero or to build it up to ${2\pi}$. We show that such a ${\pi}$-delay line can be provided by a dark soliton pinned to a potential well created by a C-shape non-resonant pump-spot. The resulting split-ring polariton condensates exhibit pronounced coherent oscillations passing periodically through clockwise and anticlockwise current states. These oscillations may persist far beyond the coherence time of polariton condensates. The qubits based on split-ring polariton condensates are expected to possess very high figures of merit that makes them a valuable alternative to superconducting qubits. \textcolor{NavyBlue}{The use of the dipole-polarized polaritons allows to control coherently the state of the qubit with the external electric field. This is shown to be one of the tools for realization of single-qubit logic operations. We propose the design of an $i$SWAP gate based on a pair of coupled polariton qubits.  To demonstrate the capacity of the polariton platform for quantum computations, we propose a protocol for the realization of the Deutsch's algorithm with polariton qubit networks.}
\end{abstract}

\maketitle

\section{INTRODUCTION}

While a tremendous progress in the development of quantum technologies is apparent, it is still unclear which material platform is the most suitable for the realisation of future quantum computers and simulators \cite{NielsenQCompQInfBook}. Among the leaders of the quest are superconducting circuits with Josephson junctions \cite{PRL112160502,Neill2018,Arute2019}, cold atoms in optical traps \cite{PRL813108,bernien2017,keesling2019}, ions \cite{SciAdv3e1601540,zhang2017,kokail2019},  purely photonic systems \cite{Brod2019},  which already provide computing facilities  on the border of capabilities of classical computing devices. The semiconductor platform legs slightly behind so far, \textcolor{NavyBlue}{ while a remarkable progress has been recently achieved \cite{Zhang2019} with spin-based quantum computing in semiconductor nanostructures \cite{Zajac2018,Yoneda2018} as well as in the creation of single-photon sources based on quantum dots \cite{Wang2019}.}
Recently, a series of papers demonstrated a high potentiality of semiconductor microcavities in the strong light-matter coupling regime for hosting ensembles of phase-locked bosonic condensates of half-light--half-matter quasiparticles: exciton polaritons (hereafter referred to as polaritons for brevity) \cite{NatMater161120,NewJPhys19125008}. It has been argued that the phase locking process in an array of polariton condensates may be used for the minimization of a classical many-body XY-Hamiltonian \cite{NewJPhys19125008}. Polariton condensates may be formed at elevated temperatures, optically controlled and mutually phase-locked on a picosecond time scale. These features constitute their main potential advantages over other material platforms for realisation of quantum simulators. On the other hand, a polariton qubit has never been convincingly demonstrated till now, and it has been argued that the dissipative nature of exciton-polaritons characterised by ultrashort radiative lifetimes would prevent their use for implementations of quantum algorithms \cite{NatMater18219}. Quite recently, an interesting proposal \cite{Ghosh2020} to build the qubit on quantized fluctuations of the resonantly driven polariton condensate in the cylinder microcavities was made.

Here we propose a quite different approach. In particular, we argue that a strong fundamental similarity of superfluid polariton flows \cite{RMP85299} and superconducting electric currents may be exploited to build a polariton analogue of the superconducting flux qubit. Superconducting flux qubits are based on superpositions of clockwise and anti-clockwise currents formed by millions of Cooper  pairs \cite{ProgTheorPhysSuppl6980,Science326113,PRL89117901}, see Figs.~\ref{fig:fig1}a,b. In order to excite the system in a superposition state, the half-quantum flux of magnetic field is passed through the superconducting circuit containing one or several Josephson junctions. The system is forced to generate a circular current to either reduce the magnetic flux to zero or to build it up to a full-quantum flux.

While electrically neutral polaritons are much less sensitive to the external magnetic  field  \cite{PRL102046407} than Cooper pairs, the circular currents of superfluid polaritons \cite{PNAS1118770} can be efficiently controlled by introducing a potential defect (a phase delay line) in a polariton ring. The defect couples counter-propagating polariton currents. This results in a formation of a two-level quantum system based on a split-ring polariton condensate. One of the efficient methods for the realization of such a defect implies pinning a dark soliton \cite{Science3321167,PRA91063625}, that is characterised by a $\pi$-jump of phase of a superfluid, to the slot in the polariton ring. The $\pi$-phase delay line embedded in a circle forces the superfluid to flow clockwise or anti-clockwise in order to either build up the phase variation along the loop to $2\pi$ or to reduce it to zero. We run numerical experiments showing the spontaneous excitation of robust  oscillations of the resulting polariton qubit state on the Bloch sphere\textcolor{NavyBlue}{, see Sec.~\ref{Sec:Dynamics}. These oscillations are caused by the continuous phase change between the qubit basis states. They are indicative of the formation of the macroscopic superposition of the qubit basis states corresponding to the symmetric and antisymmetric combinations of polariton circular currents.}

Interestingly, the predicted dephasing time of the oscillations in such a system non-resonantly pumped by a continuous-wave (cw) laser source appears to be orders of magnitude longer than the characteristic oscillation period (about 125~ps). This may result in very high figures of merit of qubits based on split-ring polariton condensates. This is because the phase gradient that governs the current states of polariton condensates is insensitive to of the overall time-dependent phase characterising the condensate as a whole object. \textcolor{NavyBlue}{Since the condensate state is continuously sustained by the external pumping, the system reaches the regime of dynamic equilibrium which is manifested by the dynamical balance between gain and dissipation. Therefore, the circular polariton currents are expected to survive as long as the  pumping is switched on \cite{Askitopoulos2020}. } 

\textcolor{NavyBlue}{
As a first step towards realization of polariton quantum networks, we propose a design of the gates performing the Pauli rotation operations which allows for the manipulation of the state of an individual qubit. {The proposed solution implies the use of the gauge field capable of lifting the degeneracy between the opposite persistent currents by additionally breaking the time-reversal symmetry. For this purpose, we propose to employ a synthetic gauge field arising from the coupling between the motion of dipole-polarized excitons and the magnetic field \cite{Lim2017}. This approach allows for the realization of a coherent control between the qubit basis states.}}

\textcolor{NavyBlue}{
For realization of a multiqubit quantum processing, we design the $i$SWAP gate based on two coupled split-ring polariton condensates. In order to demonstrate a high potentiality of semiconductor microcavities for designing of the quantum information devices, we test the system for implementability of the practical quantum protocols. 
As a benchmark test, we simulate the Deutsch's algorithm, whose realization is typically recognized as a necessary step for demonstration of the feasibility of the quantum information processing on a given material platform. This was the case of the early experiments with nuclear magnetic resonance systems~\cite{Chuang1998,jones1998} as well as with the advanced platforms, such as trapped ions~\cite{gulde2003}, superconducting circuits~\cite{DiCarlo2009}, and photonic systems~\cite{Mohseni2003,tame2007}. Implementation of the Deutsch's algorithm requires the realization of both single-qubit and two-qubit gates. The demonstration of this algorithm then turns to proof-of-concept test of the ensamble of required tools for further scaling quantum information processing systems since an arbitrary quantum algorithm can be decomposed in a sequence of single-qubit and two-qubit gates.
}

\section{RESULTS}

\subsection{The origin of the two-level quantum system}

\begin{figure}[!htb]
\centering
\vspace{0pt}
\includegraphics[width=\linewidth]{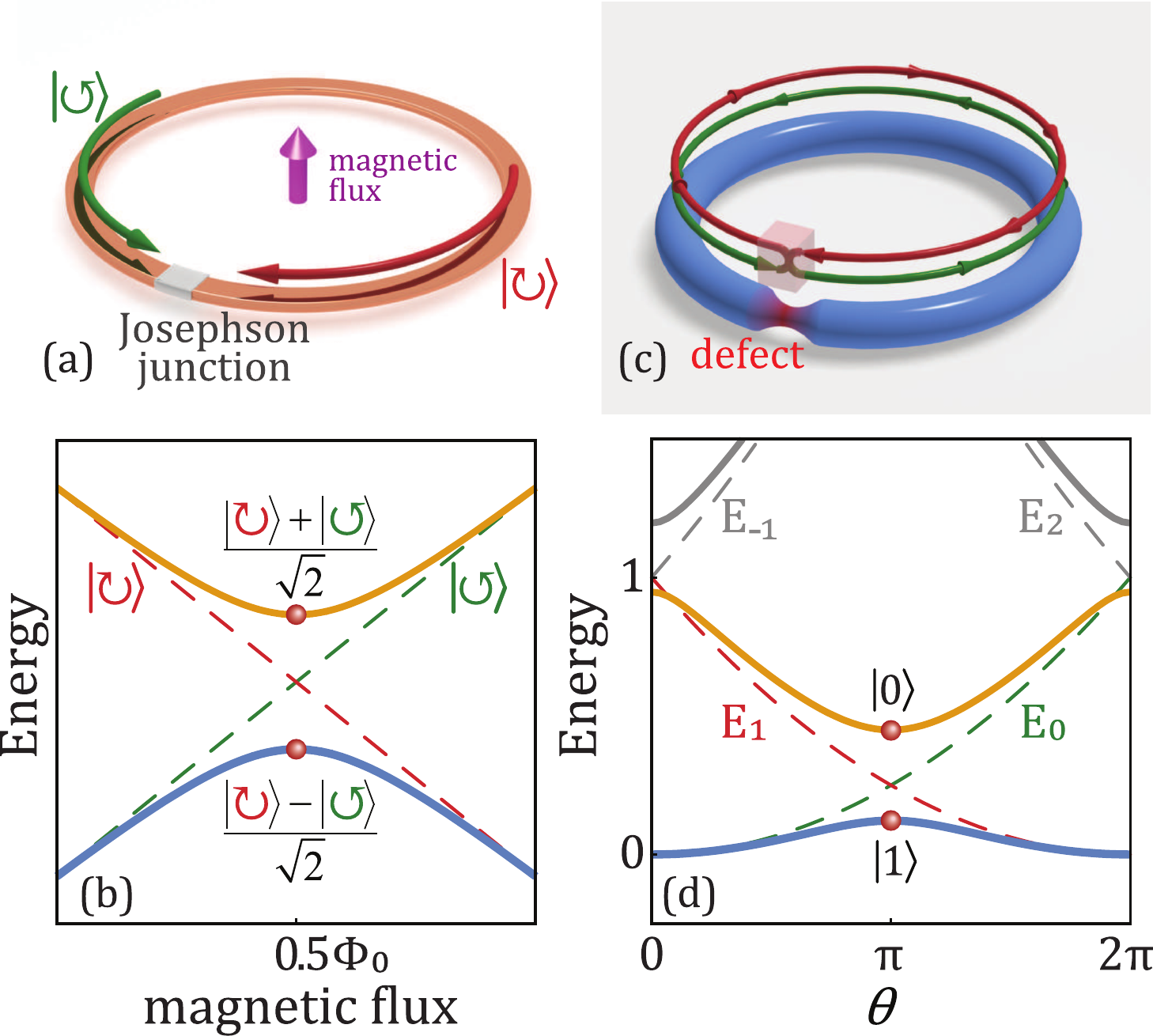}
\caption{{Comparison of a superconducting flux qubit and a split-ring polariton condensate qubit.} (a) Sketch of a flux qubit consisted of a superconductor circuit interrupted by a Josephson junction. The persistent currents generated inside the loop tend either to compensate  an external magnetic flux or build it up  to the value corresponding to the full magnetic flux quantum $\Phi_0$ ($\left|\circlearrowright \right.\rangle$ and $\left|\circlearrowleft \right. \rangle$ current states, respectively). (b) Energy levels of a flux qubit. The qubit basis is formed by symmetric and antisymmetric superpositions of the persistent current states. (c) Loop of a polariton superfluid with an embedded defect. (d) Energy diagram  of the states of the superfluid circle with different topological charges $l$ in the presence of the effective magnetic field. The energy of the state, $E_l=E_1{(\theta=0)}\left[l-\theta/2\pi\right]^2$, is measured in units of $E_1{(\theta=0)}$.
}
\label{fig:fig1}
\end{figure}

Let us consider a close circuit filled with a coherent quantum fluid.
The phase of the many-body wave function $\psi (t,s)$ of the fluid $ \varphi$ must obey the equality: $\oint _D \partial_{s} \varphi d s = 2 \pi \ell$, which is the quantisation condition for the topological invariant $\ell \in \mathbb{Z} $ also known as the winding number \cite{PRX4031052,NewJPhys19093002}.
Here $s$ is the coordinate along the circuit of a total length $D$. \textcolor{NavyBlue}{If the fluid is subjected to the gauge field, characterized by the vector potential $\mathbf{A}$,} the quantisation condition becomes $\oint _D \partial_{s} \varphi ds - \theta = 2 \pi \ell$. The phase delay $\theta= \Lambda \Phi/\hbar$ induced by the vector field is determined by its flux $\Phi$ and the constant $\Lambda$ which defines pulse rescaling rule, $\mathbf{\hat p} \rightarrow \mathbf{\hat p} - \Lambda \mathbf{A}$. The effective flux governs the energy spectrum of a two-level system based on counter-propagating currents with opposite winding numbers, as Fig.~\ref{fig:fig1}d shows.  At the particular value $\theta=\pi$, the states with $\ell=0$ and $\ell=1$ are degenerate in energy similar to the case of the superconducting flux qubit. \textcolor{NavyBlue}{These states however are different in the phase increment $\oint _D \partial_{s} \varphi ds$ which equals $\pi$ for the state with zero topological charge $\ell=0$ and $-\pi$ for the state with $\ell=1$.}

The energy gap between $\ell=0$ and $\ell=1$ states appears in the presence of a defect embedded in a circuit, see Fig.~\ref{fig:fig1}c,d. The defect causes back-scattering of the currents and mixes them. The eigenstates of this system mimic the linear superposition states realized in a superconducting flux qubit.

For electrically neutral particles the phase delay $\theta$ can be engineered either by the circular motion of the defect \cite{SciRep44298} or by exploiting the spin-orbit coupling in the presence of external magnetic fields \cite{PRL102046407}. However, the specific case of $\theta=\pi$ can be realised in a much simpler way. \textcolor{NavyBlue}{All one has to do is to embed at some point   a dark soliton state characterized by a $\pi$-jump of the phase. This can be done e.g. by pinning it with a potential defect required for localization and stabilization of the soliton state \cite{OptExpress266267,PhysRevA63.053602}.}
In the presence of a dark soliton, the current states with the phase changing by $\pi$ and $-\pi$ over the remaining part of the circuit form two superposition states  $\left|0\right.\rangle$ and $\left|1\right.\rangle$, which constitute a two-level quantum system or qubit.

\subsection{The model}

\begin{figure}
\centering
\vspace{0pt}
    \includegraphics[width=\linewidth]{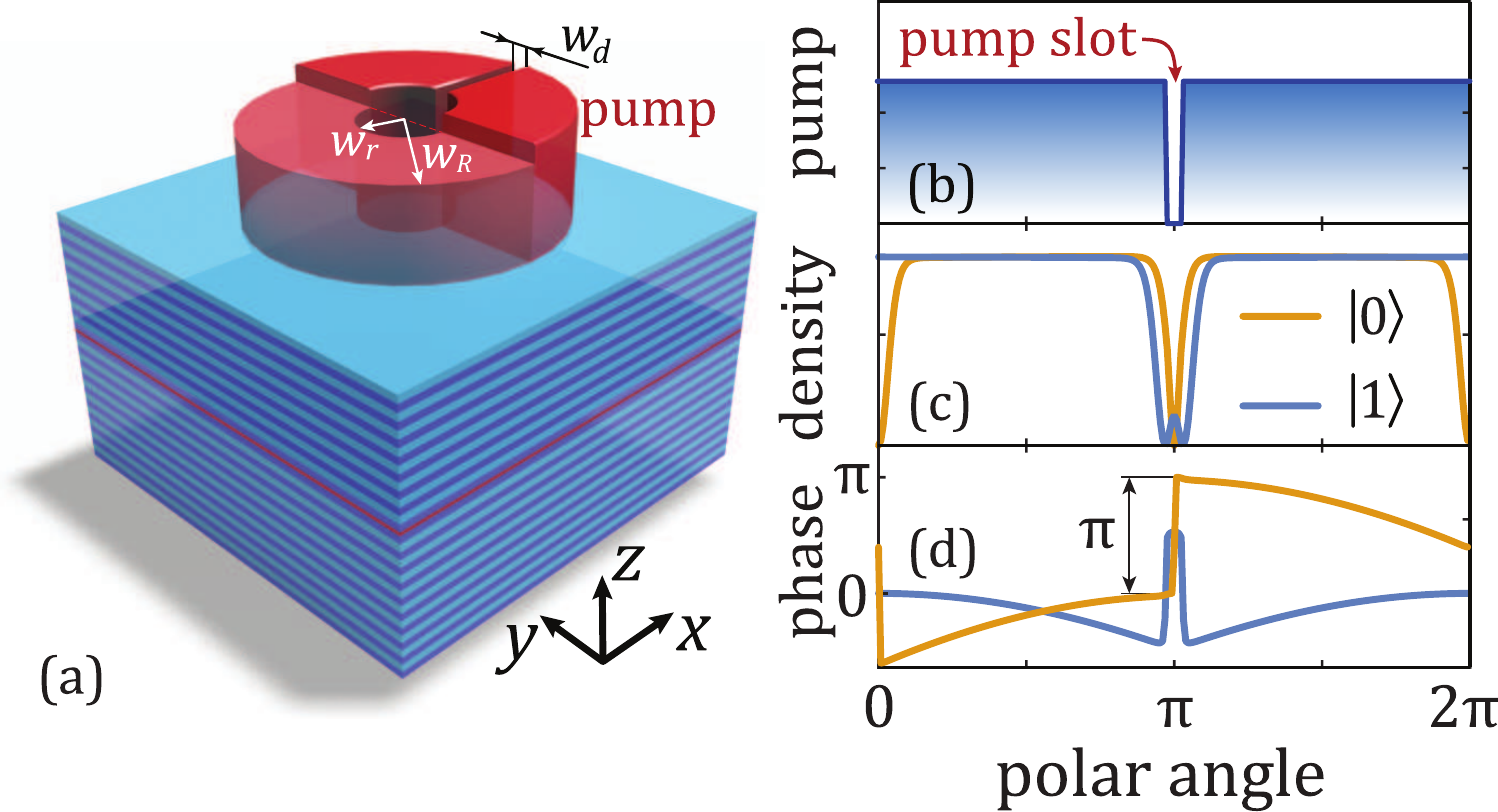}
    \caption{{The model system.} (a) the semiconductor microcavity excited by a nonresonant pump beam having the C-shape profile. Here $w_{r}=6~\mu$m and $w_{R}=20~\mu$m. The width of the slot is $w_d=2$~$\mu$m. (b)--(d) properties of the basis states of the qubit formed in a polariton superfluid ring, as predicted by the 1D model. (b) the pump distribution contains a slot which is able to pin a dark soliton manifesting itself in the density dip and the $\pi$ phase jump (panel (d)). (c) the densities and (d) the phases of the basis states $|0\rangle$ and $|1\rangle$.  (c) and (d) show the solutions of the 1D problem obtained with a pump distribution shown in (b). The pump amplitude is  $3.5\times P_{\rm th}$. }
    \label{fig:cavity_1D}
\end{figure}

We consider the system shown schematically in Fig.~\ref{fig:cavity_1D}a.
A semiconductor microcavity formed by a couple of Bragg mirrors contains an ensemble of embedded quantum wells. The strong coupling of cavity photons and quantum-well excitons results in the appearance of new eigenmodes of the structure, the exciton polaritons \cite{PRB5910830}. We assume that the polaritons are created by the non-resonant cw optical pump of a C-shape which is schematically illustrated in Fig.~\ref{fig:cavity_1D}a. Obeying the bosonic statistics, polaritons form a Bose-Einstein condensate which remains localised under the pump spot due to the finite polariton lifetime.

The dynamics of a 2D split-ring polariton condensate in a semiconductor microcavity obeys the driven-dissipative Gross-Pitaevskii equation coupled with the rate equation for the density of the exciton reservoir:
\begin{align} \label{GP1}
d \psi(\mathbf{r}) &=\left[\frac{i\hbar }{2 m^*}\nabla^{2} - \frac{i}{\hbar}\left(g_{c}|\psi(\mathbf{r})|^2 + g_rn_R(\mathbf{r}) + V(\mathbf{r})\right) \right. \nonumber \\
&+\left. \vphantom{\frac{i \hbar }{2 m^*}\nabla^{2}} \frac{1}{2}\left(R n_R(\mathbf{r}) - \gamma_c\right)\right]  \psi(\mathbf{r}) dt+dW, \nonumber \\
\frac{\partial n_R(\mathbf{r})}{\partial t} &= P(\mathbf{r}) - (\gamma_R + R|\psi(\mathbf{r})|^2)  n_R(\mathbf{r}),
\end{align}
where $\psi(\mathbf{r})$ is a polariton condensate order parameter, $n_{R}(\mathbf{r})$ is the exciton reservoir density, $m^*= 10^{-4}m_{e}$ is the effective mass of polaritons on the lower branch ($m_{e}$ is the free electron mass), $\gamma_c=1/6$~ps$^{-1}$ and  $\gamma_R = 2 \gamma_c$ are the polariton and the reservoir decay rates, respectively. $R=0.01\,\mathrm{ps}^{-1} \mu$m$^2$ is the rate of the stimulated scattering from the exciton reservoir to the polariton condensate. $g_c=6$~$\mu$eV~$\mu$m$^{2}$ and $g_r= 2 g_c$ describe the interaction of polaritons  between themselves and with the  reservoir excitons, respectively. $V(\mathbf{r})$ is the external potential. The chosen values of the parameters of the model are relevant to the experimental data \cite{Roumpos2011}.

While the classical fluctuations are taken into account in the initial polariton field, the effect of quantum fluctuations is included  by adding a complex stochastic term within the truncated Wigner approximation \cite{PRB79.165302}
\begin{align} \label{stochas}
\langle dW(\mathbf{r},t)dW(\mathbf{r}^\prime,t) \rangle &=0, \ \  \ \langle dW(\mathbf{r},t)dW(\mathbf{r},t^\prime) \rangle =0, \nonumber \\
\langle dW(\mathbf{r},t)dW^*(\mathbf{r}^\prime,t^\prime)\rangle&=\frac{dt}{2 dxdy}(R n_R+ \gamma_c) \delta_{\mathbf{r},\mathbf{r}^\prime}\delta_{t,t^\prime} .
\end{align}

The cw nonresonant pump $P(\mathbf{r})$ of C-shape, as shown in Fig.~\ref{fig:cavity_1D}a, is characterised by the spatial distribution of the intensity given by:
\begin{align}\label{pump}
P(\mathbf{r})=\begin{cases} {}
P_{0}~[1-e^{-(\frac{\mathbf{r}}{w_r})^{n}}]~[1-e^{-(\frac{2y}{w_d})^{n}}]~e^{-(\frac{\mathbf{r}}{w_R})^{n}}, \, x> 0, \\
(P_{0}-P_{1})~[1-e^{-(\frac{\mathbf{r}}{w_r})^{n}}]~e^{-(\frac{\mathbf{r}}{w_R})^{n}}, \qquad x \leq 0,
\end{cases}
\end{align}
where the pump power variation factor $P_{1}$ quantifies the depth of the step in the pump beam intensity as shown in Fig.~\ref{fig:cavity_1D}a. By additionally breaking the rotation symmetry, this step allows for manipulation of the quantum properties of the two-level system, in particular, for switching between the superposition and the pole states, as it will be demonstrated hereafter. \textcolor{NavyBlue}{ The steepness parameter is taken as $n=20$. The required profile of the pump beam can be realized using spatial light modulator \cite{PNAS1118770,NatMater161120} or by pattering the microcavity surface with an opaque mask intended to prevent excitation of polaritons beneath it.}

\textcolor{NavyBlue}{
The key ingredient of the pump  is a radial slot of the width $w_d$ where no polaritons are excited. Its function is to mix the opposite polariton currents giving rise to the energy splitting between the basis states. Besides, it favours formation of the soliton-like topological defect in the polariton condensate accompanied by the $\pi$ jump of the phase in the azimuthal direction. In particular, due to the local absence of the gain, the polariton density is typically depleted at the position of the slot. It naturally triggers formation of the dark soliton provided that the polariton-polariton interactions are strong enough to sustain it~\cite{OptExpress266267}.
Besides, the reservoir density also appears to be depleted under the slot. Since excitons repel each other, the slot is responsible for the formation of the potential well for the condensate which serves for pinning and stabilization of the soliton state.}

\textcolor{NavyBlue}{In the presence of the slot,} in addition to the persistent current states with non-zero average momenta \cite{PRA66063603}, the couple of energy non-degenerate states with zero average currents appears. These are symmetric and anti-symmetric states similar to those presented in Fig.~\ref{fig:fig1}d.  The \textcolor{NavyBlue}{typical} angular dependencies of the magnitudes and the phases of the corresponding wave functions are shown in Fig.~\ref{fig:cavity_1D}c,d.  \textcolor{NavyBlue}{The properties of these states can be conveniently analysed in the limit where the pump profile is a thin ring of a large radius, see Appendix~\ref{AppB0}.  The basis states  behave differently in the vicinity of the defect, which causes splitting in their energies. Therefore, the value of the energy gap between these states can be controlled by both shape and intensity of the pump   as it is discussed in the Appendix~\ref{AppB0}.}

\subsection{The dynamics of a split-ring condensate }\label{Sec:Dynamics}

In what follows, we shall focus on the oscillatory regime of a 2D split-ring condensate. In this regime, the condensate is initially formed in a superposition of $|0\rangle$ and $|1\rangle$ eigenstates~\cite{PRA99033830,PRB91214301}. The system exhibits long-living quantum beats whose frequency is governed by the energy splitting of $|0\rangle$ and $|1\rangle$ states. The split-ring condensate passes periodically through clockwise and anticlockwise current states. Its dynamics can be conveniently mapped to a Bloch sphere.
 At this stage we neglect by the quantum fluctuations of the condensate order parameter focusing  on its coherent dynamics.

\begin{figure}
\centering
\vspace{2pt}
\includegraphics[angle=0,width=\linewidth]{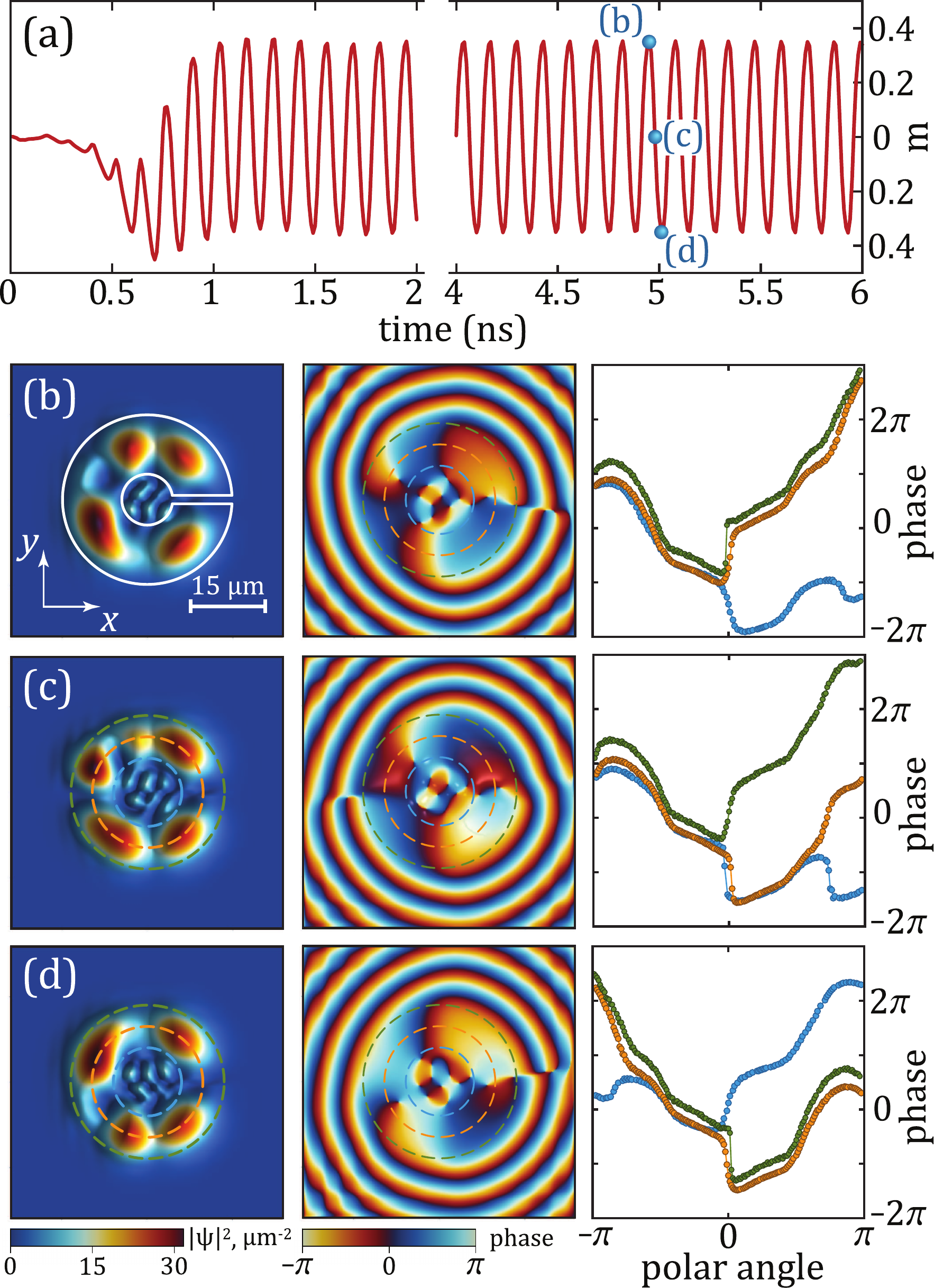}
\caption{{The temporal evolution of the split-ring polariton condensate nonresonantly pumped with the C-shape cw laser beam switched on at time zero.} (a) The temporal evolution of the mean angular momentum $m$ of the condensate. (b)--(d) The density (left column) and the phase (middle column) profiles are shown in (b,c,d) for the states with the maximum, zero and minimum values of the angular momentum, respectively. The pump shape is indicated with white contour in panel (b). The phase of the condensate as a function of the polar angle at the fixed radius $r$ is shown in the right column. Here $r=8~\mu$m, $13~\mu$m and $18~\mu$m for blue, orange and green curves, respectively.  The parameters of the pump beam used for this calculation are: $P_{0}=1.8\times P_{\rm th}$ and $P_{1}=0.1\times P_{0}$, $P_{\rm th}=\gamma_c \gamma_R/R$.}
\label{fig:fig2}
\end{figure}

Figure~\ref{fig:fig2} shows the oscillatory regime of the split-ring polariton condensate predicted by the numerical modeling of Eqs.~\eqref{GP1}. We characterize non-stationary circular current  states of the condensate by the normalized average angular momentum  $m(t) = \hbar^{-1}\left. L_z (t) \right/ N(t)$, where $L_z(t) = (-i\hbar) \int{ \psi^{*}(\mathbf{r},t)[x \partial_{y} - y \partial_{x}] \psi (\mathbf{r},t)}d\mathbf{r}$ is the actual average angular momentum and $N(t) = \int{|\psi(\mathbf{r},t)|^2} d\mathbf{r}$ is the number of polaritons in the condensate.
In contrast to the winding number $\ell$, the average angular momentum $m(t)$ continuously varies in the course of the evolution of the condensate \textcolor{NavyBlue}{and can have arbitrary real value}.
The oscillations of the polariton state in Fig.~\ref{fig:fig2}a occur between the states with the average angular momenta of $m \simeq 0.4$ and $m \simeq -0.4$.
The panels (b) and (d) show the intensity distribution (left), the phase  distribution  in the cavity plane (middle) and the angular phase distribution (right) of the polariton states with $m \simeq 0.4$ and $m \simeq -0.4$, respectively. Figure~\ref{fig:fig2}c illustrates the intermediate state of $m = 0$ visited by the system in the course of the oscillations. The full dynamics of the oscillations illustrated by Fig.~\ref{fig:fig2} is summarized in the Supplementary movie \cite{SupplMat}.

The harmonic oscillations of the angular momentum of the condensate can be considered as a fingerprint of a many-body two-level quantum system. The spectral analysis of the oscillatory dynamics of the system  reveals two sharp resonances appearing, which are split in energy by about $32$~$\mu$eV (see Appendix~\ref{AppC} and Fig.~\ref{fig:figS3}) that corresponds to the period of the oscillations seen in Fig.~\ref{fig:fig2}a.

\begin{figure}
\includegraphics[width=\linewidth]{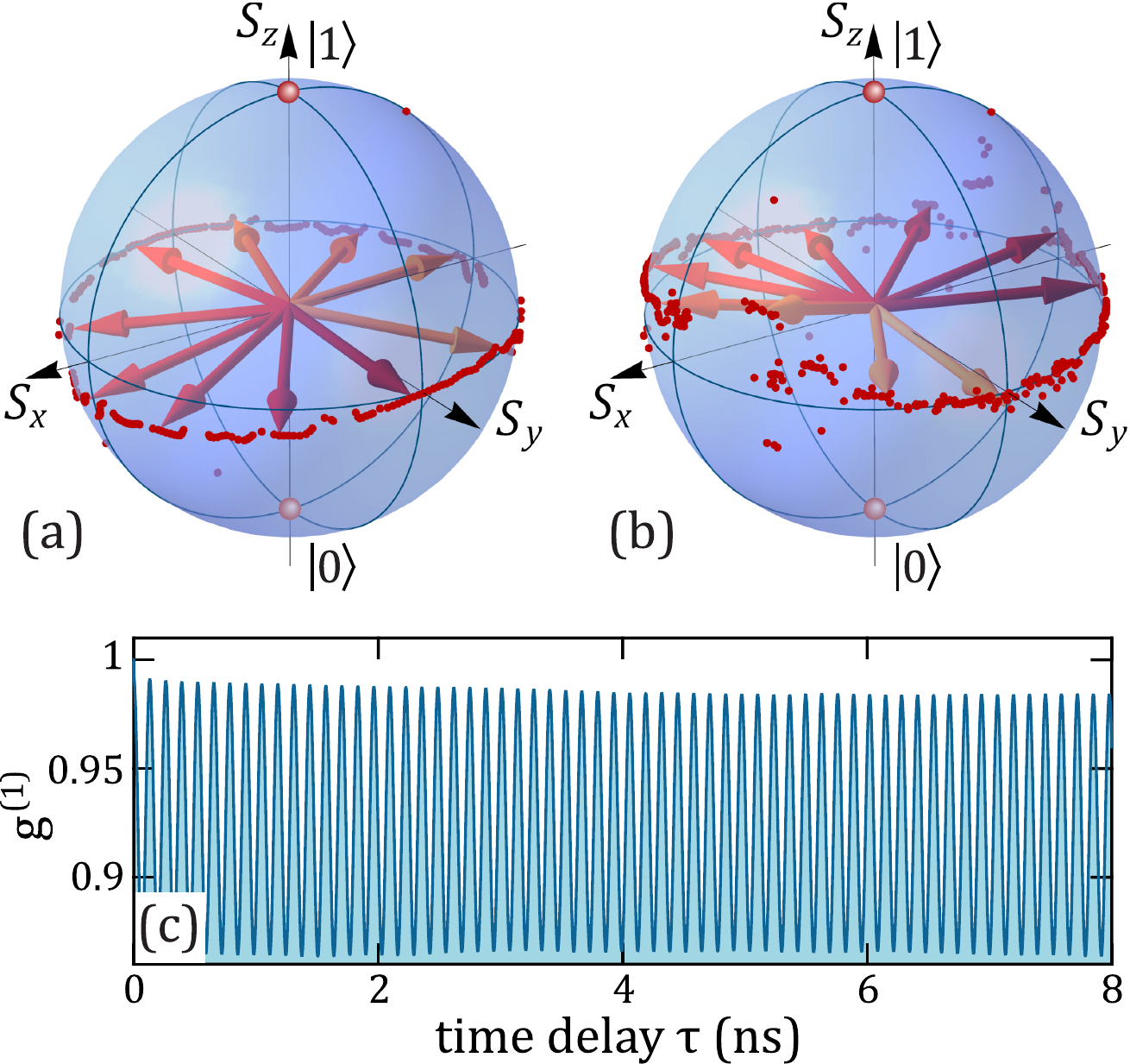}
\caption{The evolution of the quantum state of the split-ring polariton condensate on the surface of the Bloch sphere.  Four periods of oscillations are shown. (a) is computed neglecting and (b) accounting for the quantum fluctuations introduced by Eq.~(\ref{stochas}). (c) First order coherence $g^{(1)}$ as a function of time delay $\tau$ corresponding to the dynamics shown in panel (b). The decay of the envelope is caused by the presence of fluctuations. The parameters are the same as in Fig.~\ref{fig:fig2}.}
\label{fig:fig3}
\end{figure}

Figure~\ref{fig:fig3} shows the trajectory of the quantum state of the split-ring polariton condensate on the surface of the Bloch sphere based on $|0\rangle$ and $|1\rangle$ eigenstates (see details of the mapping in the Appendix \ref{AppA}).
Retaining only the classical fluctuations of the initial polariton field and neglecting the stochastic processes (Fig.~\ref{fig:fig2} and Fig.~\ref{fig:fig3}a), we obtain the stable oscillations that persist during over $10$~ns (truncation time of this numerical simulation) and correspond to a circular trajectory close to the equator of the Bloch sphere.
When including the quantum fluctuations described by Eq.~(\ref{stochas}), we find that the uncertainty in the mapping procedure to the Bloch sphere becomes larger and the trajectory of our system on the surface of the Bloch sphere looks noisy. Remarkably, the noise does not affect the stability of oscillations that persist over the whole calculation period showing no apparent decay.
\textcolor{NavyBlue}{This is because the oscillating state is continuously sustained by the pump. It can be characterized as a regime  of the dynamical equilibrium where the losses are exactly compensated by the time averaged gain.} We conclude that the harmonic oscillations in a two-level quantum system formed by a split-ring polariton condensate may persist \textcolor{NavyBlue}{as long as the pumping is on.}

 The reason of the surprising stability of oscillations is in the topological protection of circular current states in split-ring condensates. The superfluid currents are governed by the spatial distribution of the phase of the condensate. The localisation radius of the condensate ($20~\mu$m in our case) is much smaller than the coherence length in the system (over $100~\mu$m) \cite{Snoke}, which is why the coherence of superfluid polariton condensate is preserved. The coherence time of the condensate as a whole which characterises the time-dependence of the overall phase of the ring condensate does not impose limitations on the life-time of polariton currents. Note that  persistent circular currents in polariton superfluids have been experimentally observed recently \cite{Snoke,PRBB97195149}.

\begin{figure}
\centering
\includegraphics[width=\linewidth]{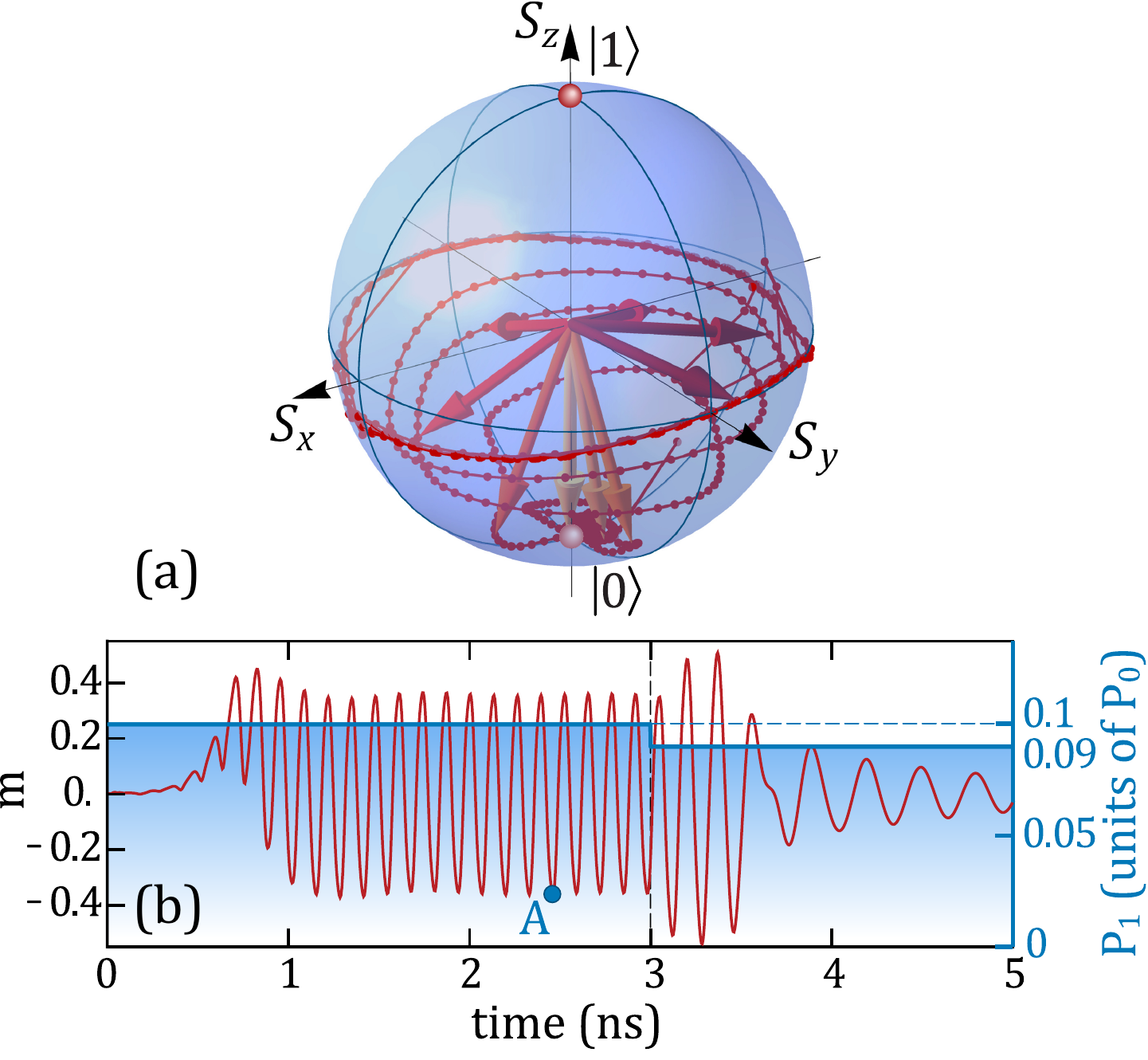}
\caption{{The oscillations of the quantum state of the split-ring condensate.}  The decay that is controlled by $P_1$: $P_{1}=0.1\times P_{0}$ while $t<3000$~ps and $P_{1}=0.09\times P_{0}$ while $t>3000$~ps. (a) the trajectory of the quantum system on the Bloch sphere shown from the point A on the time axis onward, (b) the dynamics of the angular momentum of the condensate. The switch of $P_{1}$ occurs at $t=3$~ns (vertical dashed line).}
\label{fig:fig4}
\end{figure}

\begin{figure}
\centering
\includegraphics[angle=0,width=\linewidth]{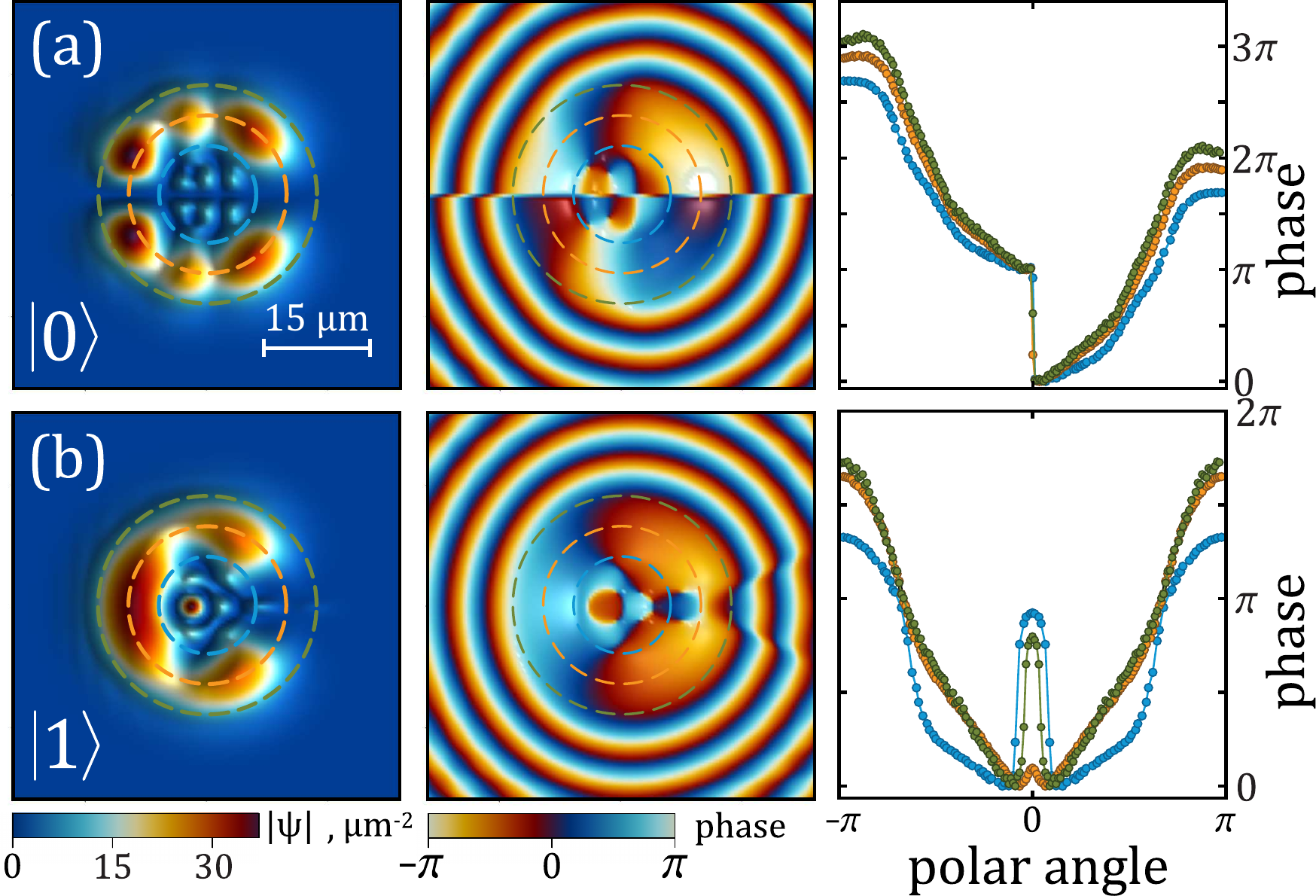}
\caption{{Two basis states of the split-ring polariton qubit corresponding to the two poles of the Bloch sphere: $| 0\rangle$ and $| 1\rangle$.} (a) $ P_{1}= 0.09\times P_{0}$ and (b) $ P_{1}= 0.03\times P_{0}$. The density and phase profile in the 2D case are shown in the left and middle panels, respectively, the angular dependencies of the phase corresponding to different fixed radii are shown with different colors in the right panels ($r=8$~$\mu$m, $13$~$\mu$m and $18$~$\mu$m are shown with blue, orange and green colors, respectively).}
\label{fig:fig5}
\end{figure}

The stable persistent oscillations of the polariton quantum system in the vicinity of the equator of the Bloch sphere are observed if the parameter $P_1$ characterising the step in the pump power distribution (see Eq.~\eqref{pump}) is chosen in the range of $[0.1,0.3]\times P_{0}$. With the decrease of $P_1$ down to $0.09\times P_{0}$, the stability of the oscillations is broken and the system exhibits a fast decay. Fig.~\ref{fig:fig4} shows the variation of the dynamics of the system resulting from the variation of the value of $P_{1}$: while $t$ is shorter than $3000$ ps, the system exhibits the same stable oscillations as those shown in Figs.~\ref{fig:fig2} and~\ref{fig:fig3}. Next, as a result of the decrease of $P_1$ from $0.1\times P_{0}$ to $0.09\times P_{0}$ at $t=3000$~ps, the fast decay of the oscillations is observed, so that, eventually, the system relaxes to one of the eigenstates, specifically, to the state $|0\rangle$ shown in Fig.~\ref{fig:fig5}a. We emphasize that the decay time of oscillations is still independent of the coherence time of the condensate in this regime. It is fully governed by $P_1$ parameter that controls the magnitude of the step potential. The trajectory of the system on the Bloch sphere that describes the decay of the oscillations has been smoothed with use of the Bezier function \cite{DunklSpecFuncBook} for clarity.

One can see in Fig.~\ref{fig:fig5}a that the wave function of the split-ring condensate is anti-symmetric with respect to the horizontal axis ($y=0$), which passes through the center of the slot. A $\pi$ phase jump appears at $y=0$. With the further decrease of $P_{1}$ down to $0.03\times P_{0}$, the system relaxes to the  basis state $|1\rangle$   shown in Fig.~\ref{fig:fig5}b.  It represents the symmetric pattern with a pair of $\pi$ phase jumps close to the horizontal axis ($y=0$).
Note that both basis states are characterised by zero average polariton flow, $m =0$, and represent 2D counterparts of the basis states shown in Fig.~\ref{fig:cavity_1D}c,d.
Besides, they are nearly perfectly orthogonal. Their orthogonality is essential for mapping the system to the Bloch sphere.

It is important to note that, in a general case, the ring condensate is not expected to relax to the lowest energy state corresponding to the upper pole of the  Bloch sphere. This is a characteristic feature of polariton lasers: out of all quantum states the system chooses one that maximises the occupation number of the condensate, but not necessarily one that is characterised by the lowest energy \cite{NatMater161120,NewJPhys19125008}. This is why incoherent processes of acoustic-phonon assistant energy relaxation are not expected to affect the dynamics of qubits based on split-ring polariton condensates.

\subsection{C-shape potentials}

Till now we were considering the polariton condensates imprinted in a planar microcavity by means of the optical pumping. Their spatial localisation was imposed by the shape of the non-resonant pump used for their excitation. An alternative way to realize split-ring polariton condensates is by using laterally confined C-shape potentials produced by chemical etching of planar cavities \cite{Mukherjee2019}. In this case, we expect a stronger confinement of polaritons and more control tools for shaping the condensates. The drawback of this system as compared to fully optically induced split-ring condensates is in its rigidity: each time, to change the geometry of an array of polariton condensates one would need to grow a new sample. We also consider the combined method of lateral confinement by using  the etched micropillars where ring condensates are formed due to the repulsion of polaritons from the exciton reservoir formed in the center of the pillar by a non-resonant optical pumping. Persistent superfluid currents of exciton-polaritons were recently observed in such structures \cite{PRBB97195149}. Shifting the pump spot from the center of the pillar yields the formation of the split-ring condensate \cite{sedov2020}. The results of numerical simulations of the harmonic oscillations in polariton condensates confined to C-shape potentials are shown in the Appendix \ref{AppB}.

\section{DISCUSSION}

\subsection{Figure of merit for split-ring polariton condensate qubits}

The simulations described above demonstrate that at certain conditions split-ring polariton condensates behave as two-level quantum systems demonstrating long-standing coherent oscillations\textcolor{NavyBlue}{, which are indicative of the continuous relative phase change between the basis states}. Considering this system as a qubit, one should be able to estimate its figure of merit
\textcolor{NavyBlue}{given by the number of logic operations that can be performed before the coherence between the two levels is lost~\cite{Yoneda2018}.
The decoherence in polariton systems is caused primarily by the combined effect of the polariton pair-scattering and quantum fluctuations of the condensate density \cite{whittaker2009}. The latter are accounted for by the noise term in \eqref{GP1} which originates from the stimulated scattering of excitons from the reservoir to the polariton condensate.
From the simulations performed, we extract the coherence time as a characteristic time of the temporal decay of the first-order coherence function $g^{(1)}$ of the condensate, shown in Fig.~\ref{fig:fig3}c. For the set of parameters used in our simulations that correspond to a conventional GaAs-based microcavity, the superposition state retains its coherence for a remarkably long time. Fitting the envelope of the the delay-dependence of the first-order coherence function with the exponential function $g^{(1)}(\tau)=\exp\left(-\tau/\tau_c\right)$, we estimate the coherence time $\tau_c$ to be over 100 nanoseconds. This is much beyond the single polariton lifetime (as short as 6 ps in our case) and even beyond the coherence time of the condensate as a whole, which may reach a few nanoseconds in optical traps  \cite{Askitopoulos2019}.
}
Thus, estimating the single logic operation time by the period of the  oscillations on the Bloch sphere that is of the order of 125 ps in our case, we end up with a figure of merit of more than \textcolor{NavyBlue}{$10^3$}, that matches those of best superconducting qubits \cite{PRL112160502}.

This high figure of merit can be achieved in a split-ring condensate because it is localised on a spot that is much smaller than the coherence length in the polariton system and because the overall phase of the condensate that is subject to a fast decoherence is fully decoupled from the superfluid phase current dynamics which defines the trajectory of the considered quantum system on a Bloch sphere. It is also important that the \textcolor{NavyBlue}{oscillating regime is sustained by the dynamical balance between gain and losses. Therefore} the energy relaxation of the condensate as a whole does not occur.

\textcolor{NavyBlue}{
One can see polariton split-ring condensates demonstrate the properties of  robust two-level systems and can be considered as qubits. In order to fully characterize the applicability of polariton qubits for quantum information processing, one should address the issues of setting the initial quantum state of a polariton qubit, coupling between different qubits, elementary logic operations and read-out of the information from a set of polariton qubits. \textcolor{NavyBlue}{In the rest of the paper we develop} the concept of quantum information processing with use of split-ring polariton qubits.
}

\textcolor{NavyBlue}{
\subsection{ Qubit state initialization  and single-qubit rotations}
}

\begin{figure*}[!htb]
\centering
\includegraphics[angle=0,width=\linewidth]{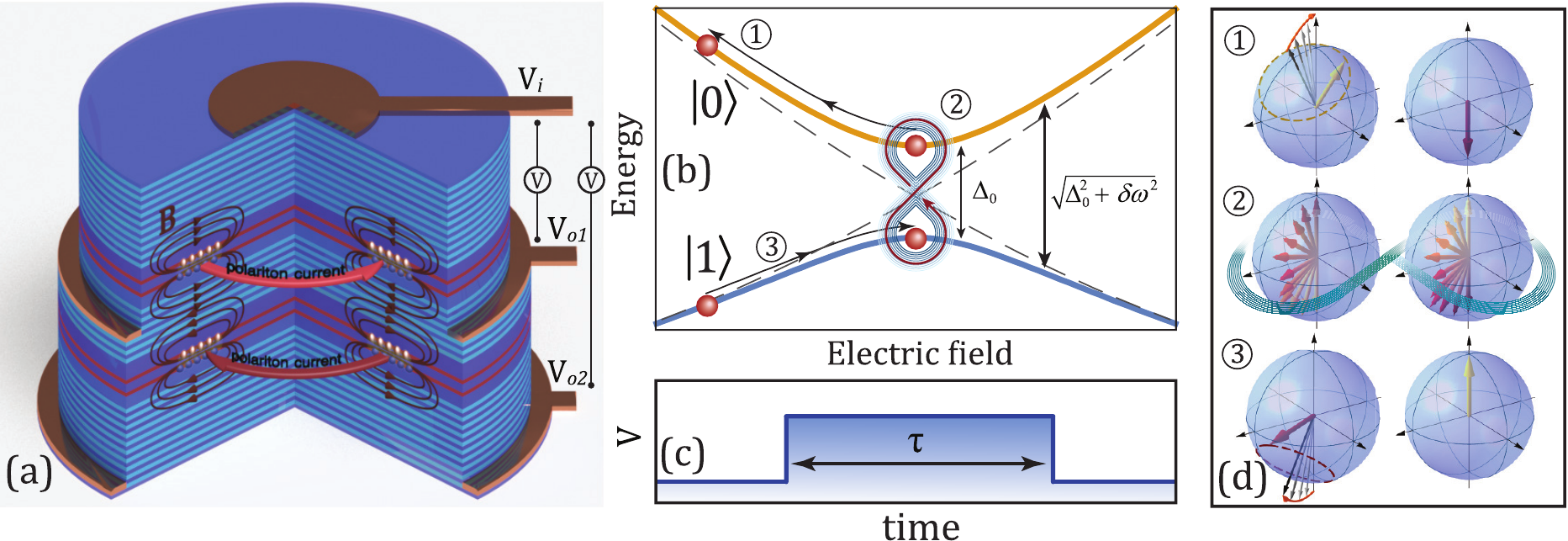}
\caption{{Two-qubit processing with split-ring polariton condensates.} (a) schematic showing  a double cavity with two coaxial dipolariton condensates.  Different positions of stop-bands of the two cavities allow for the selective non-resonant excitation and read-out of the qubits. Being detuned from the point of view of the polariton energy, two qubits are matched from the point of view of their energy gaps. Using of the spatially indirect excitons formed in coupled quantum wells endows polaritons with the permanent dipole moment aligned perpendicular to the cavity plane. A single inner ($V_i$) and a couple of outer ($V_{o1}$ and $V_{o2}$) electrodes generate the radial electric field, which in combination with the perpendicular external magnetic field introduces an artificial gauge potential required for the selective tuning of the qubits' energy gap. (b) a three-step protocol for the implementation of the $i$SWAP two-qubit gate. The energy diagram illustrates the manipulations with the energy gap of the first qubit. (c) the typical time dependence of the control electric field required for the realization of the $i$SWAP operation. (d) the schematic visualization of the Bloch vector behaviour during the implementation of the $i$SWAP gate protocol. Being out of  the resonance, the first qubit precesses about the effective magnetic field indicated be the grey arrow.}
\label{fig:fig7}
\end{figure*}

Setting a split-ring condensate into a given quantum state can be achieved with use of a resonant short pump pulse focused on a specific spot of the ring. A similar technique has been employed for setting the phase of polariton Rabi oscillations \cite{PRL113226401}. \textcolor{NavyBlue}{The state in which the qubit is excited can be controlled by tuning the depth of the pump step, while the overall phase of the condensate can be set with the use of a weak coherent pumping by a resonant laser pulse  \cite{Chestnov2019}.} 

A control of the nonresonant pump allows for implementation of single qubit properties. In particular, the total intensity of the pump affects the value of the energy gap between the basis states \textcolor{NavyBlue}{as it is discussed in the Appendix~\ref{AppB0}}. Tuning of the splitting in the two-level system is crucial for the implementation of multi-qubit operations as it allows addressing specific qubits from the quantum register.  \textcolor{NavyBlue}{Alternatively, the same goal can be achieved by lifting the degeneracy between the states characterised by the opposite angular momenta, which is equivalent to the presence of the synthetic gauge field in accordance with the energy diagram shown in Fig.~\ref{fig:fig1}d.} An existence of the similar synthetic gauge field for exciton polaritons was recently demonstrated by applying the crossed electric and magnetic fields \cite{Lim2017}. The static electric field polarizes excitons while the magnetic field causes an energy shift of the moving dipole-polarized excitons due to the magneto-Stark effect \cite{Thomas1961}. The value of the energy shift is proportional to the exciton (polariton) momentum. Therefore, this scheme is capable of synthesizing of the effective gauge potential for polaritons. In the Appendix \ref{AppE}, we discuss the specific realization of such a gauge field for a ring of dipole-polarized exciton polaritons in the presence of the radial electric field and normal to the cavity plane magnetic field. \textcolor{NavyBlue}{In particular, it is demonstrated that a single qubit dynamics is governed by the following Hamiltonian written in the truncated two-level basis:
\begin{equation}\label{Ham1qb}
\hat H = \frac{\Delta_0}{2}\sigma_z + \frac{\delta\omega}{2} \sigma_x,
\end{equation}
where $\Delta_0$ is the energy of the basis states splitting, $\delta\omega$ is the bias parameter corresponding to the gauge field-induced splitting between the circular polariton flows with opposite momenta.}

\textcolor{NavyBlue}{The presence of the synthetic gauge field is crucial for implementing the quantum logic with split-ring qubits. Manipulation of the  persistent current states splitting parameter $\delta\omega$ paves the way for realization of the  most of the single-qubit quantum gates. Here we demonstrate the performance of the operations of elementary rotation of the qubit state vector also known as Pauli rotation gates. }

\textcolor{NavyBlue}{As discussed in detail in the Appendix \ref{AppE}, we assume that the gauge field responsible for this splitting is controlled by the external voltage applied to the ring electrodes, Fig~\ref{fig:fig7}a. \textit{Pauli $X$- and $Y$-gates} (in what follows denoted as $\mathcal{R}_x$ and $\mathcal{R}_y$) require a resonant periodic driving of the qubit. This is achievable with the AC electric pulses, which  generate a time-periodic radial electric field polarizing excitons in the cavity plane. The driving force applied in resonantly to the qubit, triggers the periodic beats of population  between the basis states $|0\rangle$ and $|1\rangle$ that are similar to atomic Rabi oscillations. In the Bloch sphere representation it corresponds to the rotation about the axes lying in the equatorial plane. The $\mathcal{R}_x$ and $\mathcal{R}_y$ rotations are distinguishable by the choice of the relative phase between the qubit and the driving force, see Appendix~\ref{AppF} for  details. Note that the AC driving of the qubit provides a convenient tool for the  coherent control between the basis states. This technique can be used for the realization of  Rabi or Ramsey interferometry experiments  for  verification of the predicted long-lived coherence of split-ring polariton qubits. }

\textcolor{NavyBlue}{
The \textit{$Z$-gate} $\mathcal{R}_z(\theta)$ operation rotating the state vector by an angle $\theta$ about the $z$ axis is equivalent to the application of the time delay $\theta/\Delta_0$ between two successive single-qubit operations.  Indeed, according to Hamiltonian \eqref{Ham1qb} written in the  frame of reference of the unbiased qubit, at $\delta\omega = 0$, the state rotates with the frequency $\Delta_0$ about the polar axis of the Bloch sphere.}

\textcolor{NavyBlue}{The Pauli rotations should be considered as the  basis operations since their combinations can be used for the realization of new gates. In the Appendix~\ref{AppF} we demonstrate the synthesis of two successive unitary rotations resulting in implementation of the Hadamard gate, which allows for switching between the superposition and the basis states.} 

\textcolor{NavyBlue}{
\subsection{Coupling between split-ring polariton qubits}
}
\textcolor{NavyBlue}{
The interaction between split-ring condensates is a key ingredient for the realization of a scalable material platform for quantum computing.  The efficient mechanism of the coherent coupling between polariton condensates is their exchange by polaritons propagating in the plane of a microcavity.
This simple coupling scheme proved efficient for pairing of nearest neighbours in an array of polariton condensates employed in XY-simulators \cite{NewJPhys19125008}. In order to achieve coupling of distant condensates one can use one-dimensional optical waveguides imprinted lithographically in the microcavity plane \cite{PRL101016402}.}

\textcolor{NavyBlue}{Here we propose an alternative approach that allows for
fabrication of compact two-qubit gates.} We demonstrate that the coupling between split-ring qubits is possible due to the magnetic field-mediated interaction between dipole-polarized condensates.  If  polaritons are polarized in the direction normal to the cavity plane, their circular flow induces separated counter propagating electric currents comprised of the carriers of opposite signs. The magnetic field generated by these currents affects another condensate excited nearby, which picks up the magnetic field in the same fashion as an inductive coil in an electric transformer. The interaction between the qubits, as expected, is maximized in the stacked geometry realized with the double cavity (Fig.~\ref{fig:fig7}a) made with  use of the diluted magnetic semiconductors such as CdMnTe to increase the magnetic susceptibility of the system.

With the truncation to the two-level basis, the Hamiltonian of the magnetic field-mediated qubit-qubit interaction reads (see Appendix \ref{AppD})
\begin{equation}\label{Hint}
\hat{H}_{\rm int}= g \sigma_x^{(1)} \otimes \sigma_x^{(2)},
\end{equation}
where the Pauli $x$-operator $\sigma_x^{({\rm i})}$ acts on $\rm i^ {th}$ qubit spinor $\bm{\psi}^{({\rm i})} = \left( | 0 \rangle ,  | 1 \rangle \right)^T$; $g$ is the interaction strength associated with the mutual inductance of the ring condensates of dipolar polaritons.
\textcolor{NavyBlue}{The magnetic field-mediated interaction \eqref{Hint} represents a versatile tool for quantum information processing. 
Eq.~\eqref{Hint}  resembles the Heisenberg Hamiltonian describing  the exchange interaction between two spins. The interaction of this type
 is known for its efficiency for quantum computing \cite{DiVincenzo2000,Schuch2003}. In particular, it is naturally suitable for   the state swapping operations as it will be demonstrated in the next subsection.}

\textcolor{NavyBlue}{Another important advantage of the inductive coupling is that its strength can be controlled by the external bias. 
This can be demonstrated recalling that in the diagonal basis the state of the individual qubit precess about the $z$-axis according to
}
\begin{equation}
\hat{H}^{({\rm i})}= \Delta_i \sigma_z^{({\rm i})}/2,
\end{equation}
\textcolor{NavyBlue}{where $\Delta_i=\sqrt{\Delta_{0i}^2 + \delta\omega_i^2}$ is the $\rm i^ {th}$ qubit eigenfrequency in the presence of the bias}. Therefore, in the  rotating wave approximation, which implies neglecting of the rapidly oscillating terms, the interaction Hamiltonian reads
\begin{equation}\label{iSWAPHam}
H_{\rm int}=  g \left(e^{i\delta t/\hbar} \sigma_+^{(1)} \otimes \sigma_-^{(2)} + e^{-i\delta t/\hbar} \sigma_-^{(1)} \otimes \sigma_+^{(2)} \right),
\end{equation}
where $\sigma_\pm = \sigma_x \pm i\sigma_y $ are the rising and the lowering operators which switch the qubit state between two poles; $\delta=\Delta_1 - \Delta_2$ is an energy detuning between the qubits. At  resonance, $\Delta_1 = \Delta_2$, the interaction (\ref{iSWAPHam}) acts to exchange the quantum states between the qubits. The two-qubit state then oscillates with the frequency $2g/\hbar$ between the states $|0\rangle \otimes |1\rangle$ and $i|1\rangle \otimes |0\rangle$, see Appendix~\ref{AppG}. On the contrary, out of resonance  the interaction appears to be effectively suppressed due to the time averaging of the interaction Hamiltonian (\ref{iSWAPHam}). \textcolor{NavyBlue}{This
remarkable effect can be exploited for the realization of the on-demand switching of qubit-qubit interaction by tuning the corresponding eigenfrequencies with the electric field}

\textcolor{NavyBlue}{
\subsection{Realization of the $i$SWAP gate with coupled split-ring polariton condensates}
}
 \textcolor{NavyBlue}{As a practically important example of  application of the inductive coupling of the split ring qubits, we propose the protocol of implementation of the $i$SWAP two-qubit gate, as Fig.~\ref{fig:fig7} illustrates.} 
The $i$SWAP operation is designed to permute the states of the two qubits with the addition of $\pi/2$ phase difference. The protocol consists of three steps. In the first step, two qubits are prepared in a state characterised by the mismatch in their energy gaps. The energy mismatch guarantees that the magnetic field-mediated interaction is suppressed. Then we bring the qubits into resonance adiabatically tuning the pump intensity or using the external electric field\textcolor{NavyBlue}{, see Fig.~\ref{fig:fig7}b}. This triggers the recurring exchange of the states between the qubits at the frequency $2g/\hbar$. The oscillations are interrupted after the time period $\tau = \hbar\pi/2g$ by detuning the qubits out of resonance. This constitutes the third step of the protocol. As a result of this gate operation, the condensates exchange their quantum states with the acquisition of the relative $\pi/2$ phase shift.

\textcolor{NavyBlue}{
\subsection{Qubit state read-out}
}
Finally, the read-out of a quantum state of the qubit can be done combining the time- and spatially-resolved photoluminescence and interferometry measurements \cite{sedov2020}. Note that this is a ``weak measurement'' method that does not fully destroy the measured quantum state, while it perturbs it to some extent. Conceptually, in a similar way, a SQUID-based read-out perturbs but does not fully destroy the quantum state of a superconducting flux qubit \cite{PRL112160502,NewJPhys18055016}. The proposed optical read-out technique is currently being used for studies of XY-simulators based on an array of exciton-polariton condensates \cite{NewJPhys19125008}.

\textcolor{NavyBlue}{
\subsection{Implementation of the Deutsch's algorithm with coupled split-ring polariton condensates}
}

\textcolor{NavyBlue}{The use of an electric field for implementation of both single and two-qubit gate operations allows for concatenation  of these gates into the logic circuits suitable for realization of practical quantum algorithms. As an illustrative example of a high potentiality of the polariton split-ring qubit computational platform, we simulate here the implementation of the quantum Deutsch's algorithm \cite{Deutsch1992}.}

\textcolor{NavyBlue}{
Being one of the first oracle-based quantum protocols, the  Deutsch's algorithm was designed to demonstrate an advantage of quantum computing in a task of determining whether a coin is fair or fake. Clearly, in a classical domain, one needs at least two queries to cope with this task. The quantum Deutsch's algorithm, in contract, requires one examination step only.
}

\begin{figure}
\includegraphics[width=\linewidth]{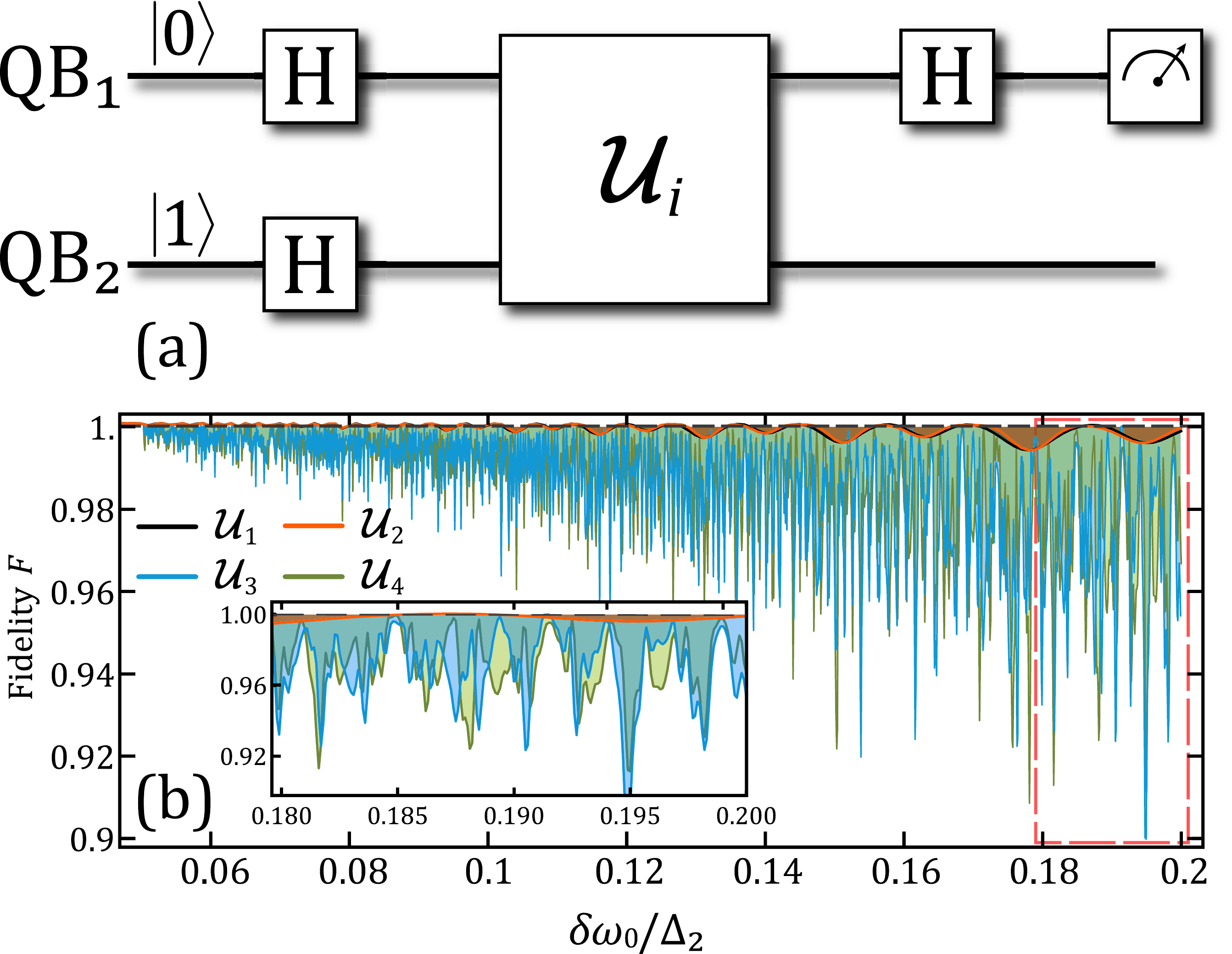}
\caption{Implementation of the Deutsch's algorithm. (a) The logical circuit. $\mathbf{H}$ denotes the Hadamard transform. The output state of the first qubit is measured to detect whether the coin (function) is fair (balanced) or fake (constant). (b) The total fidelity of the algorithm simulated at different levels of the  control pulse amplitude $\delta\omega_0=\rm{max}\left[\delta\omega_1(t)\right] = \rm{max}\left[\delta\omega_2(t)\right]$. Four realizations of the algorithm corresponding to the oracles $\mathcal{U}_1$, $\mathcal{U}_2$, $\mathcal{U}_3$, $\mathcal{U}_4$ are indicated by the black, orange, blue and green colors respectively. An example of the electric pulse sequence required for implementation of the $\mathcal{U}_3$ gate is illustrated in the Appendix~\ref{AppH}.
}\label{Fig.DeutschAlgorithm}
\end{figure}

\textcolor{NavyBlue}{
The work of the algorithm can be illustrated using a set of four functions $f_i$, where $i\in \{1,2,3,4\}$, which map the input bit $x$ (being either $0$ or $1$) onto the output bit $y$, i.e. $y=f_i(x)\in\{0,1\}$. 
The functions $f_i$ can be divided into two classes: the constant functions
\begin{subequations}
\begin{equation}\label{Eq.constF}
f_1(x) = 0  \, \, \,  {\rm{and}} \, \, \, f_2(x) = 1,
\end{equation}
whose output does not depend on the input state, and the balanced functions
\begin{equation}\label{Eq.balanF}
f_3(x) =  x,  \, \, \,  {\rm{and}} \, \, \, f_4(x) = {\rm{NOT}}(x),
\end{equation}
\end{subequations}
where ${\rm{NOT}}$ is the classical bit-flip operation.
It is assumed that one gets some function $f_i$ with unknown uniformly distributed random $i\in \{1,2,3,4\}$ as a black box. For classical bits, it is possible to characterize the given function as constant or balanced by measuring its output twice: tacking consequently $0$ and $1$ states as the input. The quantum Deutsch's algorithm copes with this problem within only one query.
It represents each of the functions $f_i$  by the corresponding two-qubit quantum gate  $\mathcal{U}_i$ (see Appendix \ref{AppH} and Table~\ref{Table1} therein) which plays the role of an oracle.
Incorporating these gates into the logical circuit shown in Fig.~\ref{Fig.DeutschAlgorithm}a, one can determine whether the function is balanced or constant by only a single call of the oracle. The answer to the problem appears to be encoded in the output state of the first qubit: The qubit is in the state $|1\rangle$ for balanced functions and in the state $|0\rangle$ in the opposite case.
}

\textcolor{NavyBlue}{
The gates $\mathcal{U}_i$ required for realization of the Deutsch's algorithm are represented by the identity and the bit-flip operations for the constant functions \eqref{Eq.constF} and by the controlled NOT-gate  and the zero-controlled NOT-gate for balanced ones \eqref{Eq.balanF}. All these functions  can be synthesized from the single-qubit Pauli rotations and the $i$SWAP gates, see Appendices~\ref{AppG} and~\ref{AppH} for the gate identities.}

\textcolor{NavyBlue}{For the practical realization of the Deutsch's algorithm with the split-ring condensates one needs to prepare two polariton qubits designed with the mismatch of their energy gaps, say $\Delta_2 > \Delta_1$. The $i$SWAP-gate has to be implemented by applying the DC voltage pulse  on the first electrode, see Fig.~\ref{fig:fig7}, which governs the bias of the qubit \#1. The pulse amplitude is chosen such as to bring the first qubit into the resonance with the second, ${\rm max}\left[\delta\omega_{i{\rm SWAP}}\right] =\sqrt{\Delta_2^2 - \Delta_1^2}$. The Hadamard gates are realized with two AC pulses as it is described in the Appendix~\ref{AppF}.}

\textcolor{NavyBlue}{
Fig.~\ref{Fig.DeutschAlgorithm} illustrates the total fidelity of the algorithm measured  as $F=\left|\langle \Psi|_{{\rm out}, i}\, \mathcal{U}_{{\rm{D}}i} |\Psi \rangle_{\rm in} \right|^2$, where $|\Psi\rangle$ is a four-component vector of the two-qubit system with $ |\Psi \rangle_{\rm in} = |01\rangle$ being the input state and $|\Psi\rangle_{{\rm out}, i}$ being the simulated output of the total algorithm realized with the gate $\mathcal{U}_i$; $\mathcal{U}_{{\rm{D}}i}$ denote the transformation matrices of the ideal circuit with the gate $\mathcal{U}_i$. 
For the details of simulations see Appendix~\ref{AppH}.
The main factor reducing the algorithm fidelity is the magnitude of the control pulse parametrized by the bias amplitude $\delta\omega_{0} ={\rm max}\left[\delta\omega\right(t)]$ in Fig.~\ref{Fig.DeutschAlgorithm}. Indeed, the accuracy of the Pauli rotation operations (specifically, the $\mathcal{R}_z$) is limited by the validity of the rotating wave approximation, which typically assumes that the driving force is weak. For simple realizations such as those involving constant gates $\mathcal{U}_1$ and $\mathcal{U}_2$, the fidelity precesses close to $F=1$ as the ratio $\delta\omega_0/\Delta_2$ grows. The sequence of local rotations applied during the implementation of the more complicated realization of the quantum algorithm results in the frequent beating of the total fidelity, see the curves describing $\mathcal{U}_3$ and $\mathcal{U}_4$ quantum gates in Fig.~\ref{Fig.DeutschAlgorithm}. Notably, the average error always remains within few percent. It justifies a high potential of the developed microcavity platform for quantum computing.
}

\section{CONCLUSIONS}
We have demonstrated that a coherent many-body quantum system represented by a bosonic condensate of exciton-polaritons placed in a split-ring geometry sustains stable and long-living oscillations between two circular current states. The polariton system qualitatively reproduces the behaviour of a superconducting flux qubit. In a remarkable similarity to the flux qubit, in the considered split-ring polariton condensate a two-level quantum system is formed by superposition states of clockwise and anticlockwise circular currents. The size of the system is much less than the coherence length of a polariton condensate, which is why superfluid polariton currents are well preserved. This ensures a high figure of merit for qubits based on split-ring polariton condensates. We propose \textcolor{NavyBlue}{a specific approach for implementation of single- and two-qubit logic operations which implies application of the external the electric field for the coherent control of the qubit state. Namely, we design an $i$SWAP two-qubit gate, which is known for its capability of producing the entangling gates required for a universal quantum computing.}
The present analysis paves the way to the realisation of a new semiconductor platform for quantum information processing. The evident advantages of the considered quantum system are in its high scalability,  high operation temperature, ultrafast logic operations and potential integrability with classical semiconductor based nano-electronic devices.

\begin{acknowledgments}
This work is supported by the Westlake University (Project No.~041020100118) and Jilin University (the Fundamental Research Funds for the Central Universities). XM and SS acknowledge the Deutsche Forschungsgemeinschaft (DFG) through the TRR142 (grant No.~231447078, project A04) and Heisenberg program (Grant No.~270619725).  A.K. acknowledges fruitful discussions with Boris Altshuler and Yuriy Rubo.
\end{acknowledgments}

\appendix

\section{Mapping the dynamics on the Bloch sphere and calculation of the mean angular momentum} \label{AppA}

For a qubit based on a two-level system formed by the states $|0 \rangle$ and $|1 \rangle$, any state $|\psi \rangle$ on the surface of the Bloch sphere can be represented as a linear combination of two basis states: $|\psi \rangle=\alpha  |0 \rangle + \beta |1 \rangle$, where the normalization condition  $|\alpha|^2+|\beta|^2=1$ is implied. The circular current states can be represented as: $\left|\circlearrowright\right. \rangle=e^{- i \frac{\pi}{4}}\left({|0 \rangle +~i|1\rangle}\right)/{\sqrt{2}}$, $\left|\circlearrowleft\right. \rangle =e^{i \frac{\pi}{4}}\left({|0 \rangle -~i|1 \rangle}\right)/{\sqrt{2}}$.
For simplicity, we associate $|m \simeq \pm0.4\rangle$ (points corresponding to the plots (b) and (d) in Fig.~\ref{fig:fig2}) with $\left|\circlearrowleft\right. \rangle$ and $\left|\circlearrowright \right. \rangle$ and choose these states as the basis for our numerical fitting procedure. In this way, we obtain:
\begin{align} \label{fit}
|\psi \rangle &=\alpha_0  |m \simeq 0.4 \rangle + \beta_0 |m \simeq -0.4 \rangle.
\end{align}
The quantum state visited by the condensate can be characterized by a pseudo-spin vector $\mathbf{\hat{S}}=S_x \mathbf{\hat{e}}_x+S_{y} \mathbf{\hat{e}}_y+S_{z}\mathbf{\hat{e}}_z$ whose components are defined as: $S_x=\frac{1}{2}(\alpha \beta^{*}+\alpha^{*}\beta)$, $S_y=\frac{i}{2}(\alpha^{*}\beta -\alpha \beta^{*})$, $S_z=\frac{1}{2}(|\beta|^{2}-|\alpha|^{2})$. The mapping of the condensate dynamics to the Bloch sphere is realised using the method of Maximum Inherit Optimization.

We note that the current direction at the position of the slot is different for the inner part ($r<r_{0}$) and the outer part ($r>r_{0}$) of the polariton condensate, as shown in the right panels of Fig.~\ref{fig:fig2}b-d ($r_0=8$~$\mu$m). Fig.~\ref{fig:fig2}a shows the angular momentum for the outer part of the polariton condensate that manifests a pronounced superfluid phase current.

\textcolor{NavyBlue}{
\section{Engineering energy gap with the C-shape optical pump}\label{AppB0}
}

\textcolor{NavyBlue}{ {Although the C-shape pump shown in Fig.~\ref{fig:cavity_1D} has a complicated profile, it is the radial slot which plays the most essential role. It governs the energy splitting of the basis states, which is a key characteristic of the qubit. In this section we consider how the incoherent pump can be used to control the value of the energy gap. }
Properties of the basis states $|0\rangle$ and $|1\rangle$  are  analysed with the simplified one-dimension equivalent of the 2D model \eqref{GP1}. This assumption is relevant to the limit of a thin ring with a large mean radius $R_0$~\cite{Mukherjee2019}. The transformation to the 1D model is performed with a substitution $\psi(\mathbf{r},t)=\varPhi(\rho)\bar{\psi}(\phi,t)$ and $n_R(\mathbf{r},t) = N_R(\rho)\bar{n}_R(\phi,t)$, where $\varPhi(\rho)$ and $N_R(\rho)$ account for the radial distribution of the condensate and the reservoir, respectively, $\rho$ and $\phi$ are polar coordinates. After integrating out the radial dependence and neglecting the stochastic term, one reduces Eqs.~\eqref{GP1} to the two coupled 1D ordinary differential equations for $\bar{\psi}(\phi,t)$ and $\bar{n}_R(\phi,t)$ with $\nabla^2 \rightarrow \partial_{\phi\phi}$ and $m^* \rightarrow R_0^2m^*$. }

\textcolor{NavyBlue}{
Then tacking $\partial \bar{n_R}/\partial t=0$ and $\bar{\psi}(\phi,t) = \bar{\psi}_0(\phi)\exp({-i\mu t})$ we solve the obtained stationary problem for $\bar{\psi}_0(\phi)$ and $\bar{n_R}(\phi)$ iteratively using the Newton-Raphson algorithm. This method yields simultaneously the azimuthal dependence of the wave function and the corresponding eigenenergy $\hbar\mu$. The typical shapes of the basis state  wave functions  are displayed in Fig.~\ref{fig:cavity_1D}c,d. Note that the solutions predicted with the model of the reduced dimensionality demonstrate a good agreement with those shown in Fig.~\ref{fig:fig5}, which were obtained from the dynamical simulations performed using of the full 2D model. It justifies the validity of the approach employed.
}

\begin{figure}
\includegraphics[width=\linewidth]{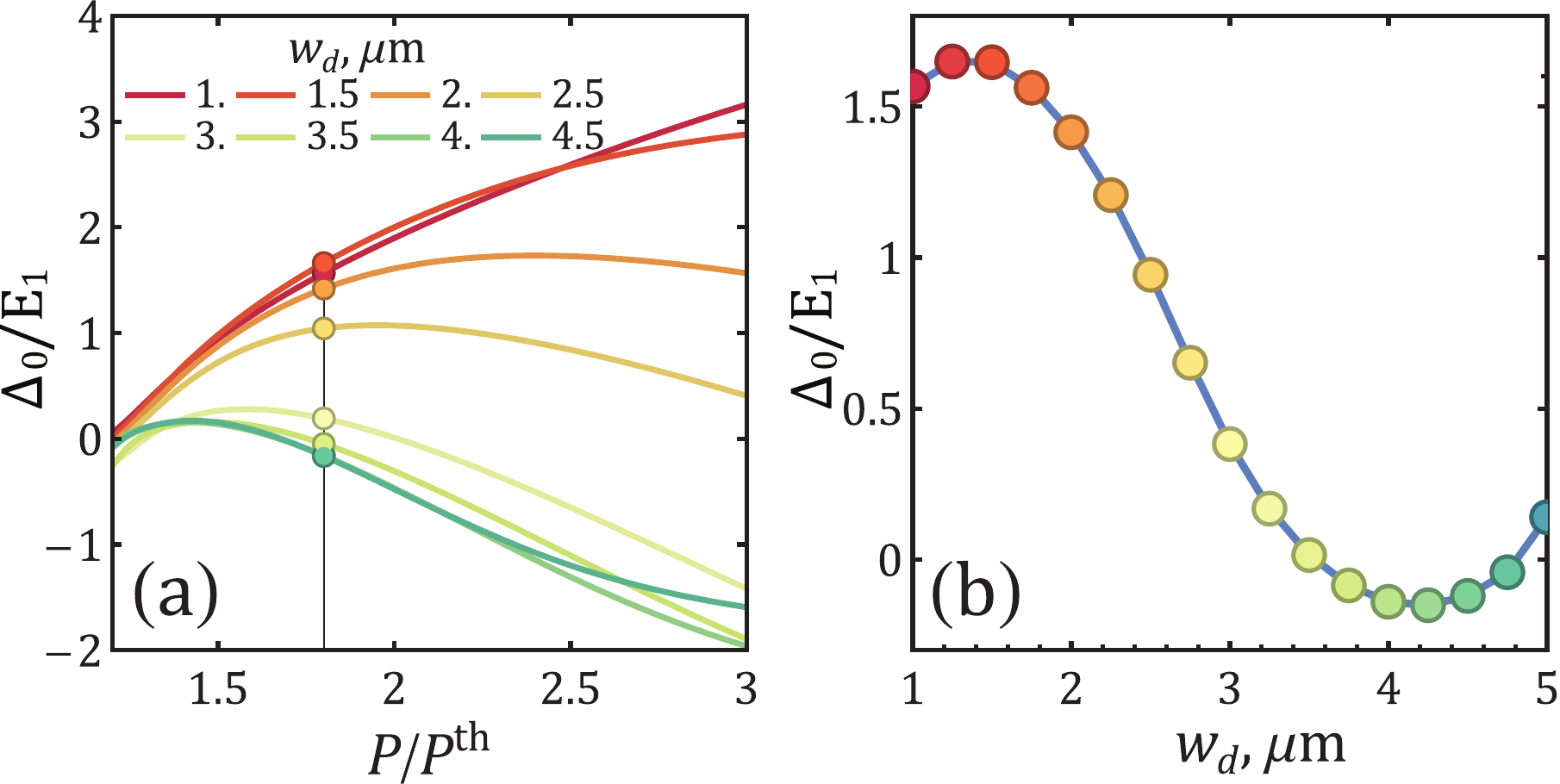}
\caption{\textcolor{NavyBlue}{Properties of the energy gap $\Delta_0$ between the qubit basis states measured in the units of energy of the circular current with a unit vorticity $E_1=\hbar^2/2m^{*}$. The mean radius of the pump ring is $R_0=15$~$\mu$m. Other parameters necessary for calculations are inherit from the full model \eqref{GP1}.}}\label{Fig.Gap}
\end{figure}

\textcolor{NavyBlue}{
The dependencies of the energy splitting between the states  $|0\rangle$ and $|1\rangle$ on the pump parameters are shown in Fig.~\ref{Fig.Gap}. We focus on  the impact of the slot width $w_d$ and the top-hat region amplitude $P_0$. For simplicity, we do not account for the pump step, working with the pump profile illustrated in Fig.~\ref{fig:cavity_1D}b. Note that the gap $\Delta_0$ demonstrates a nonmonotonic behaviour as a function of the total pump power, see panel (a). Although the particular $\Delta_0(P)$-behaviour essentially depends on the slot width, the gap typically decreases close to the condensation threshold $P^{\rm th}$ and saturates at the pump power being of about several times the threshold. The complicated dependence of the energy gap on the pump intensity indicates the significant role of the polariton-polariton interactions, which are responsible for the dependence of the energy of the condensate  on its population. The variation of the slot width provides a fine tuning of the energy gap as it is demonstrated in Fig.~\ref{Fig.Gap}b. For the considered parameters, the optimal value of the gap is achieved with the slot  width being about few micrometers and the pump power being not too far above the threshold.
}

\textcolor{NavyBlue}{
We see that  the incoherent pump provides an efficient mechanism for engineering of the properties of the split-ring qubit.
The tuning of the qubit splitting is a promising tool for realization of the single-qubit control. However, the manipulation speed appears to be limited by the response time of the spatial light modulator.
An alternative approach which implies the use of the synthetic gauge field, as discussed in the Appendix~\ref{AppE}, appears to be suitable for establishing the coherent control between the basis states of the qubit. Besides, this method is characterized by the fast response, which is governed by the timescale of the exciton dipole-polarization formation, and therefore is more attractive from the point of view of applications in quantum computing.
}

\section{Half-quantum currents trapped in C-shape potentials}\label{AppB}
\begin{figure} [!htbp]
\centering
\includegraphics[width=\linewidth]{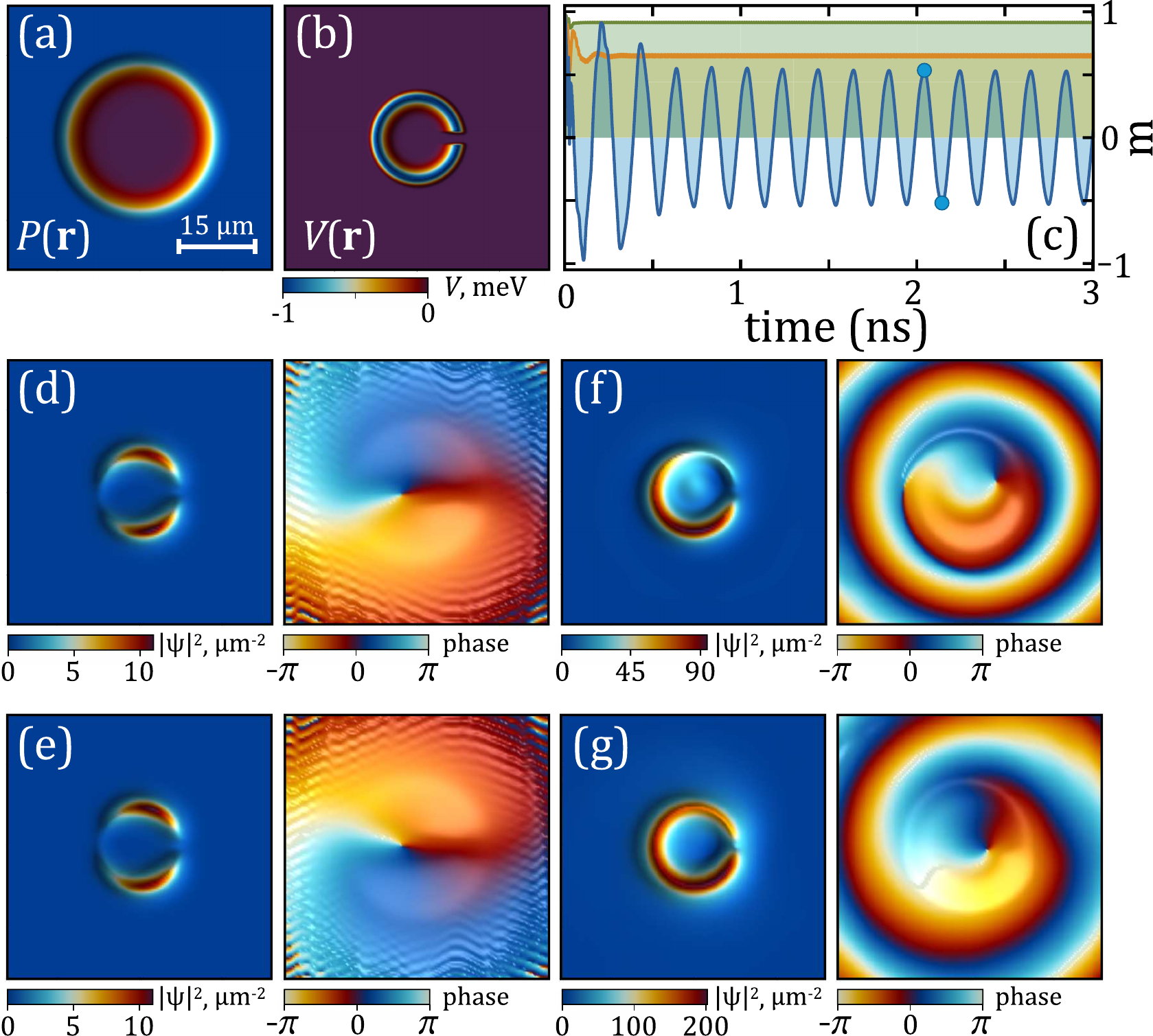}
\caption{{Circular superfluid currents confined to a narrow C-shape external potential.} (a) Pump profile with a flat top. (b) Narrow C-shape potential well with the potential depth of $1$~meV. (c) Time evolution of the angular momentum of the condensate for different pump intensities: $P_0=1.3\times P_{\rm th}$ (blue line), $P_0=3.3\times P_{\rm th}$ (orange line), and $P_0=6.6\times P_{\rm th}$ (green line). Density (middle row) and phase (bottom row) distributions of the condensate at different pump intensities: (d,e) $P_0=1.3\times P_{\rm th}$, (f) $P_0=3.3\times P_{\rm th}$, and (g) $P_0=6.6\times P_{\rm th}$. (d) and (e) correspond to the blue points in (c). $P_{\rm th}$ is the condensation threshold. Here, the parameters are: $\gamma_{\rm c}=0.3$~ps$^{-1}$, $m^{*}=2\times10^{-5}m_{\rm e}$, $g_c=3\times10^{-3}$~meV~$\mu$m$^{2}$.}\label{narrowCpotential}
\end{figure}
Circular superfluid currents with fractional angular momenta demonstrating persistent coherent oscillations can also be found in the case of a condensate confined to a C-shape external potential created e.g. by etching of a planar microcavity sample. In order to describe the system in this case we introduce the additional stationary potential $V(\textbf{r})$, see Eq.~\eqref{GP1}. We consider the non-resonant excitation of the system by a broad pump as illustrated in Fig.~\ref{narrowCpotential}. The considered potential contains a narrow barrier, which is different from the optically induced potential distribution shown in Fig.~\ref{fig:cavity_1D}. In the main text, the slot in the pump-ring corresponds to a potential well for the polariton condensate. Under the excitation by a non-resonant broad pump with a relatively low intensity ($P_0=1.3\times P_{\rm th}$), an oscillating state with its angular momentum varying between $m\simeq\pm$0.5 is obtained, as shown in Figs. \ref{narrowCpotential}c-e.

\begin{figure} [htbp]
\centering
\includegraphics[width=0.5\linewidth]{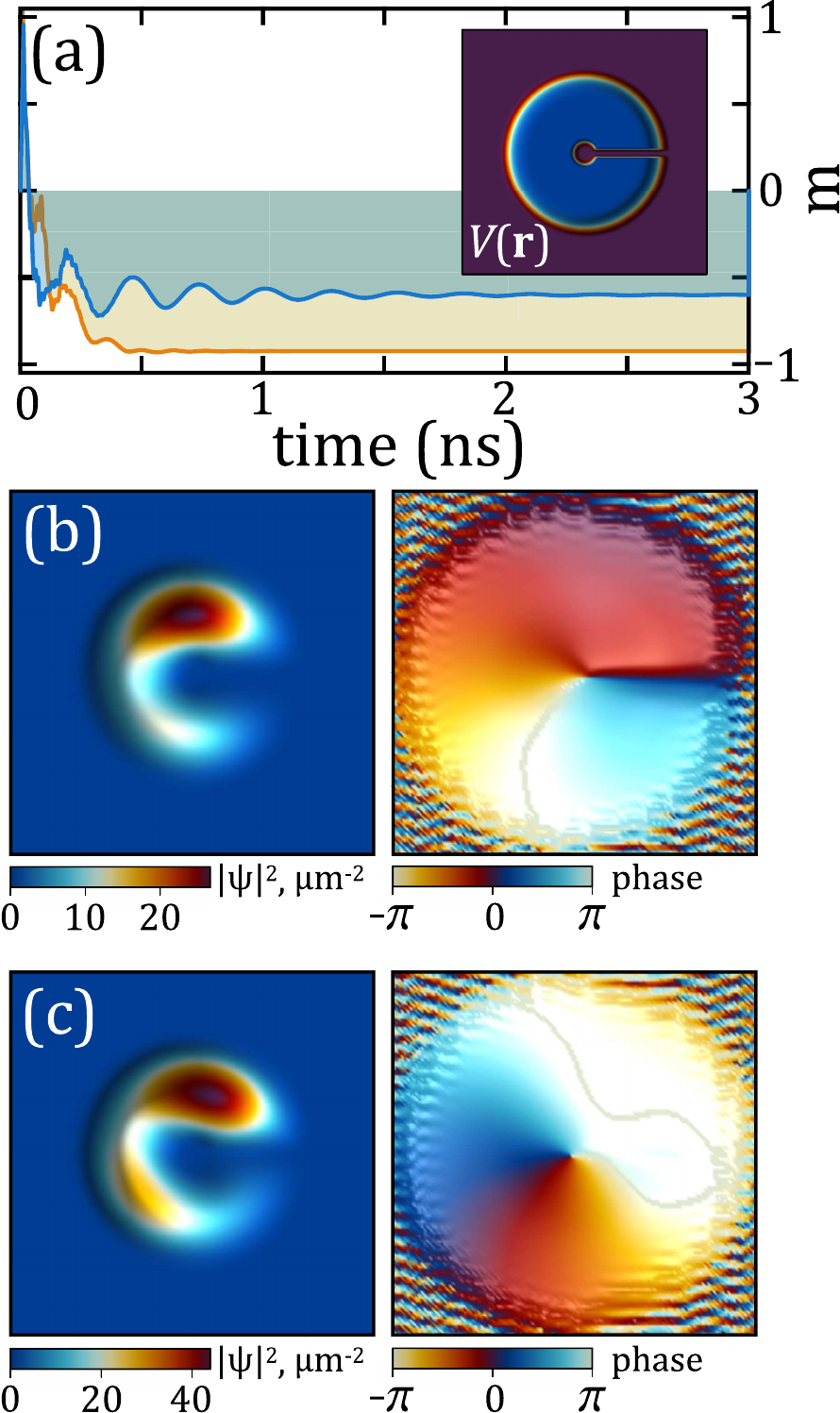}
\caption{{Circular superfluid currents confined to a wide C-shape external potential.} (a) Time evolution of the angular momentum for different pump intensities: $P_0=2 \times P_{\rm th}$ (blue color) and $P_0=2.6\times P_{\rm th}$ (orange color). The inset shows the broad C-shape potential well of the depth of $1$~meV. Density (right column) and phase (left column) distributions of condensates at different pump intensities: (b) $P_0=2\times P_{\rm th}$ and (c) $P_0=2.6\times P_{\rm th}$.}\label{broadCpotential}
\end{figure}

The tunneling of polaritons through the narrow barrier mimics the Josephson dynamics. It is important to underline that Josephson oscillations between two condensates observed in Ref.~\cite{PRL105.120403} decay on the timescale of the coherence time of polariton condensates because of the decoherence between two condensates. In contrast, in the ring geometry we work with a single condensate. The fluctuations of its overall phase do not affect the phase difference between its parts situated to the right and left sides of the potential barrier, which is why the oscillations persist on a much longer time-scale in our case. We estimate that the decoherence time in our system scales exponentially with the ratio of the coherence length to the diameter of the ring condensate. It may exceed several tens of nanoseconds in realistic GaAs-based microcavities.

As the pump intensity increases, the system achieves a steady state regime where the normalised angular momentum is fractional, as shown in Fig.~\ref{narrowCpotential}f. Note that this state is different from those in Fig.~\ref{fig:fig2}b,d, since there is no clear $\pi$ phase jump at the potential barrier here. Making the pump intensity much stronger, one can see clearly the tunneling of polaritons under the potential barrier, see Fig.~\ref{narrowCpotential}g. The tunneling ensures a smooth phase variation in this case, so that the angular momentum of the condensate approaches $m=1$, see Fig.~\ref{narrowCpotential}c.

We note that the shape of the external potential strongly influences the spatial distribution of the polariton density in the condensate. If the potential width is increased (inset to Fig.~\ref{broadCpotential}a), the same broad pump as in Fig.~\ref{narrowCpotential}a can create a C-shape solution with the angular momentum $m\simeq\pm0.5$ (Fig.~\ref{broadCpotential}a), and a clear $\pi$ phase jump~\cite{PhysRevA63.053602} is observed at the potential barrier as shown in Fig.~\ref{broadCpotential}b. This solution is very similar to that of Fig.~\ref{fig:fig2}. As in the previous case, whilst the pump intensity increases, the phase difference between both sides of the potential barrier changes from $\pi$ to $0$, leading to the formation of the current state with an integer angular momentum, see Figs.~\ref{broadCpotential}a,c.

\section{Spectral analysis of the oscillations in a split-ring polariton condensate}\label{AppC}

\begin{figure}
\centering
\includegraphics[width=0.8\linewidth]{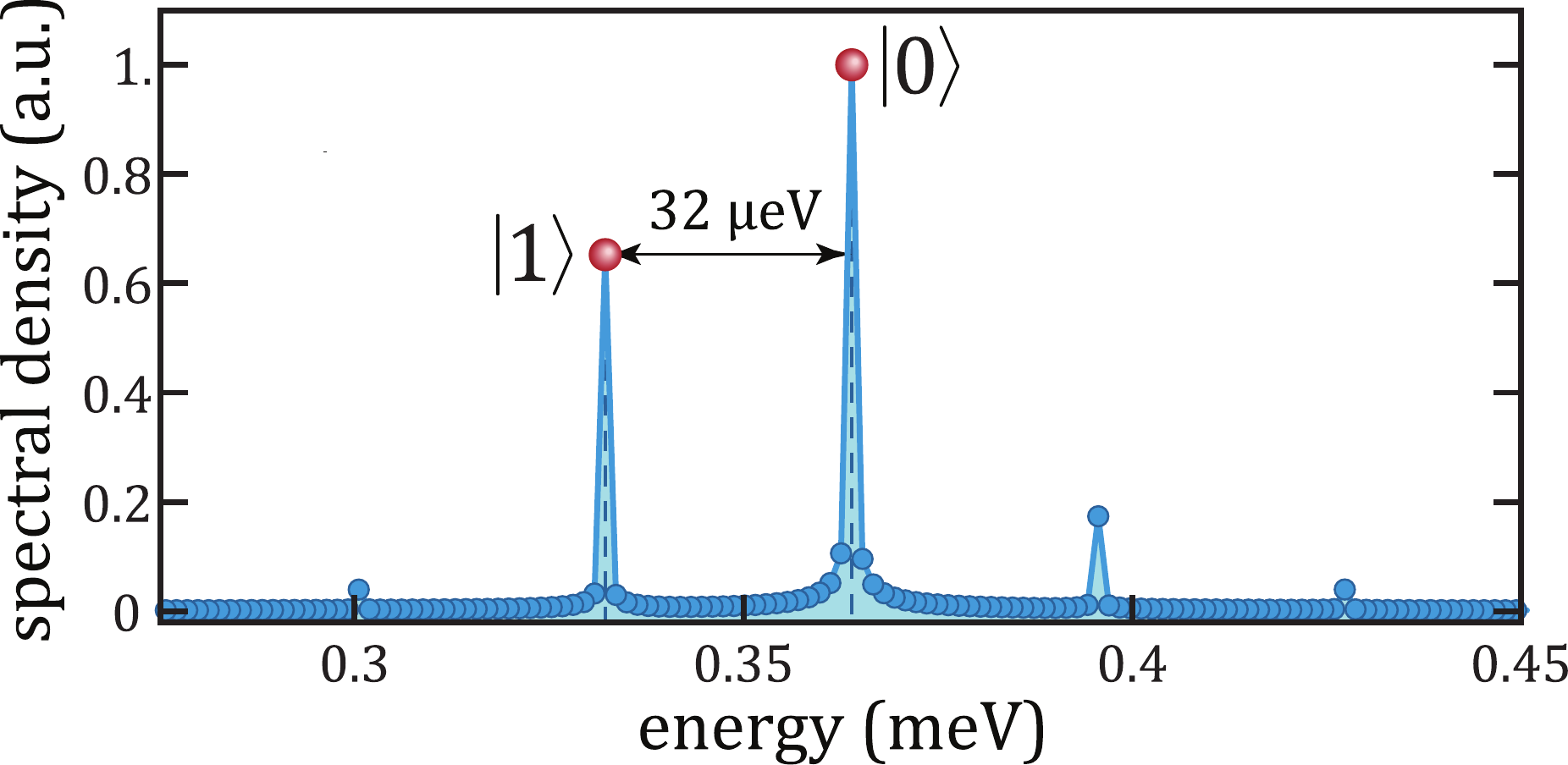}
\caption{{The spectrum of the split-ring condensate.} The energy spectrum of the split-ring condensate in the oscillating regime corresponding to Fig.~3\ref{fig:fig2}.}
\label{fig:figS3}
\end{figure}

The strong evidence of the two-level nature of the oscillating split-ring polariton condensate is provided by its energy spectrum. The Fourier spectrum of the dynamics of the condensate wave-function  (see Fig.~\ref{fig:fig2}) $S(\omega)=\iint \psi(t,\mathbf{r})e^{-i\omega t}dt d^2\mathbf{r}$ is shown in Fig.~\ref{fig:figS3}.  Note that besides the pair of peaks corresponding to the eigenstates of the system $\left|0\right.\rangle$ and $\left|1\right.\rangle$ shown in Fig.~\ref{fig:fig5}, the comb of the side-band peaks of attenuated intensities also appears in agreement with the recent predictions \cite{PRL114.193901,PRB101.085302}. These satellites result from the nonlinear processes in the driven-dissipative polariton system and are indicative of the deviation of the considered split-ring condensate from an ideal two-level linear quantum system. The intensities of the side peaks increase as the pump power increases and the interactions between counter-rotating polariton currents become important.

\section{Magnetic field-assisted coupling between circular polariton currents}\label{AppD}

Polaritons are composed by the electrically neutral excitons and photons. Therefore, in the first order approximation over the ratio on the exciton Bohr radius to the magnetic length their motion is typically uncoupled from the magnetic field in the sense that moving polaritons experience the same energy shift from the magnetic field as the polaritons at rest. In particular, it is true for the polaritons formed by the ground-state excitons, which possess no electric dipole moment. However, it is not correct for those polaritons, which appear due to the strong coupling between the light and the dipole-polarized excitons. The stationary dipole moment is characteristic excitons excited e.g. in coupled asymmetric quantum wells, where the electron and the hole are localized in spatially separated layers. This particular case is illustrated in Fig.~\ref{fig:fig7}. Besides, the exciton dipole moment can be induced and, what is more important, controlled by applying an external electric field. The latter case has a crucial importance for tuning the qubit energy gap, as we show in the next  section.

Now let us consider a ring-shaped condensate of dipole-polarized polaritons.
In what follows we address the properties of the excitonic component only and neglect the spin-dependent effects. However, since the light-matter interaction in the strong coupling regime implies  conservation of momentum, the exciton and the polariton momenta are identical. Therefore, the coupling of the dipolar exciton motion with the magnetic field is naturally imprinted on  polariton properties with the  scaling factor given by the squared excitonic Hopfield coefficient $C_x$.

According to \cite{JETP42.449} the motion of the exciton center of mass is coupled to its internal structure in the presence of the external magnetic field $\mathbf{B}$. This coupling is reflected in the appearance of the extra term in the excitonic Hamiltonian:
\begin{equation}\label{HamExdB}
\hat{H}_{\rm ex}\propto \frac{e\left[\mathbf{B} \times  \mathbf{d}\right]}{M_{\rm ex}}\left( -i\hbar \bm{\nabla} \right),
\end{equation}
where $e$ is the elementary charge, $M_{\rm ex}$ is the exciton effective mass associated with its motion as a whole particle. The bold symbols are used for the  vectorial variables. In particular, $\hbar \mathbf{k} = -i\hbar \bm{\nabla}$ is the exciton momentum, $\mathbf{d}$ is the electric dipole moment.
The Hamiltonian \eqref{HamExdB} implies that the coupling is maximized provided that the electric dipole, magnetic field and the exciton momentum are mutually orthogonal. Therefore, for the polaritons polarized perpendicularly to the microcavity plane, $\mathbf{d}=\mathbf{d}_z$, and rotating about the $z$-axis, $\mathbf{k}=\mathbf{k}_\varphi$, it is the radial component $\mathbf{B}_{\rho}$ of the magnetic field that affects the polariton motion. Here $z$, $\rho$ and $\varphi$ are the cylindrical axes.
Besides in this particular configuration  the radial magnetic field modifies the exciton dispersion shifting its minimum out of the zero-momentum state.

The effect, which is in some sense reciprocal to the one described above, is the generation of the magnetic field by a ring of rotating dipolar exciton polaritons. Indeed, the motion of the exciton is equivalent to the two counterpropagating currents of the oppositely charged electron and hole. Since the carriers are separated in space, the magnetic fields they generate do not  compensate each other. The resultant field is maximized in the interlayer region where the radial component  $\mathbf{B}_{\rho}$ is the dominating one.
For the magnetic field generated by the polariton fluid with the unit vorticity (whose winding number is one), its out-of-plane radial component is
\begin{equation}\label{Brho}
B_\rho =  \frac{\mu_0e\hbar C_x^2 n_{\rm pol}}{2M_{\rm ex}\rho^2} d,
\end{equation}
where $ n_{\rm pol}$ is the polariton density, $C_x^2$ is the squared exciton Hopfield coefficient that quantifies a contribution from the excitonic fraction. Note that the expression (\ref{Brho}) is valid in the region where $\rho \gg h \gg d$, where $h$ is the distance from the quantum well plane. The more general expression can be found in \cite{PRB56.10355}.

According to the previous discussion, the field (\ref{Brho}) generates the polariton current in the co-axial polariton condensate excited in the adjacent microcavity located at the distance $h$. This effect can be  taken into account in the driven-dissipative Gross-Pitaevskii equation~\eqref{GP1}. The Hamiltonian (\ref{HamExdB}) yields additional $\mathbf{k}$-dependent terms in the relevant Schr\"odinger equation for the exciton wave function. They  can be accounted for in Eq.~\eqref{GP1} using the coupled oscillator model. Here we omit the rigorous derivation of the relevant expression leaving it for a more detailed study (see also \cite{Lim2017}).

Besides, for the sake of simplicity we consider the limiting case of the homogeneous polariton density disregarding the dark soliton perturbations in the vicinity of the defect. With this approximation Eq.~\eqref{GP1} for the wave function of the first condensate $\Psi^{(1)}$ acquires an additional term which describes a coupling with the second condensate associated with the wave function $\Psi^{(2)}$:
\begin{equation} \label{S3}
i\hbar \frac{\partial \Psi^{(1)}}{\partial t} \propto \frac{\mu_0 e^2 d^2 \hbar}{2M_{\rm ex}^2 R_0^3} C_{x1}^2 C_{x2}^2 n_{\rm pol}^{(2)} \,\, \hat{L}_z \Psi^{(1)},
\end{equation}
where we used a non-stochastic form of the Eq.~\eqref{GP1} and skipped all terms but  one which arises from the Hamiltonian (\ref{HamExdB}). The values  $C_{x1}$ and $C_{x2}$ are assigned to the excitonic Hopfield coefficients characterizing the first and the second condensate, correspondingly.
In (\ref{S3}) we defined the angular momentum operator as $\hat{L}_z= -i \hbar R_0 \partial / \partial\varphi$. This assumption is valid for the quasi-one-dimensional case, i.e. for the condensate localized on a thin ring with the radius $R_0$. Then, defining the average angular momentum per unit area of the second condensate as $L_z^{(2)}= S^{-1}\int_S \left({\Psi^{(2)}}\right)^* \hat{L}_z \Psi^{(2)} d\mathbf{r} = \hbar m n_{\rm pol}^{(2)}$, where $m$ is a mean angular momentum defined in the main text, we obtain:
\begin{equation}\label{S4}
i\hbar \frac{\partial \Psi^{(1)}}{\partial t} \propto g L_z^{(2)} \hat{L}_z \Psi^{(1)},
\end{equation}
where  $g= \frac{\mu_0e^2 d^2}{2M_{\rm ex}^2 R_0^3}C_{x1}^2 C_{x2}^2$.

Associating the condensate angular momentum with the $x$-component of the  pseudospin state $\bm{\psi}$ written in the basis of the non-rotating states $|0\rangle$ and $|1\rangle$, we can write the truncated Hamiltonian which is capable of describing the evolution of the coupled split-ring qubits:
\begin{equation}\label{2qbHam}
\hat{H}_{qq} = \sum_{\rm i=1,2} \Delta_{ i} \sigma_z^{(\rm i )}/2 + g \sigma_x^{(1)} \otimes \sigma_x^{(2)},
\end{equation}
where $\Delta_{i}$ defines the energy gap between the basis states of the $\rm i^{th}$ qubit and the $\otimes$ sign emphasises the tensor product. Note that the coupling strength $g$ can be associated with the effective mutual inductance using the analogy with the superconducting flux qubit systems.

\section{Tuning the energy gap with the artificial gauge field}\label{AppE}

Implementation of the most of the two-qubit quantum gates implies a sequence of operations with the individual qubits. In particular, for the realization of the $i$SWAP gate protocol proposed in the main text, one needs a tool for  tuning the energy gap between the qubit basis states. In the case of superconducting flux qubits, this is easily achievable by varying the external magnetic flux (see Fig.~\ref{fig:fig1}c). For the exciton polaritons such a strategy is also possible. However it requires   the synthetic gauge field which would act on the electrically neutral polaritons. As it was demonstrated in \cite{Lim2017}, such an artificial field arises from the coupling between the motion of a polarized exciton and the magnetic field given by Eq.~(\ref{HamExdB}). Here we explain how this approach can be used for the tuning of the energy gaps in the split-ring polariton condensates.

We consider the set-up shown in Fig.~\ref{fig:fig7}a. An external electric field $\mathbf{E}$ has a radial component which induces the exciton polarization characterized by the dipole moment $\mathbf{d}_\rho = \alpha \mathbf{E}_\rho$, where $\alpha$ is the exciton polarizability. The external magnetic field is uniform and it is directed perpendicularly to the microcavity plane, $\mathbf{B}=\mathbf{B}_z$. This configuration guarantees  the mutual orthogonality of the magnetic field, the dipole moment and the azimuthal polariton momentum $\hbar \mathbf{k}_\varphi$.
In this case, the artificial gauge potential $\mathbf{A}$, which defines the pulse rescaling rule $\mathbf{\hat p} \rightarrow \mathbf{\hat p} - \mathbf{A}$, reads
\begin{equation}
\mathbf{A}= \mathbf{A}_\varphi=e \left[\mathbf{B}_z \times  \mathbf{d}_\rho\right] C_x^2 m^*/M_{\rm ex},
\end{equation}
where $m^*$ is the effective mass of polaritons on the lower dispersion branch.
In the uniform polariton density approximation that was assumed in the previous section, and in the defect-free case the splitting between the states whose winding numbers differ by~1~is
\begin{equation}\label{Eq.bias}
\delta\omega(E_\rho)= \frac{2e \hbar C_x^2 B_z}{R_0 M_{\rm ex}} d_\rho(E_\rho).
\end{equation}
\textcolor{NavyBlue}{Equation \eqref{Eq.bias} explicitly shows the electric field-dependence of the energy bias parameter $\delta\omega$.}

Note that the action of the effective field is equivalent to the condensate rotation about the $z$-axis. It lifts the energy degeneracy between the clockwise and anticlockwise currents. Therefore, in the truncated basis this rotation is equivalent to the appearance of the additional term in the single qubit Hamiltonian:
\begin{equation}
\hat{H}_q = \frac{\Delta_0}{2} \sigma_z + \frac{\delta\omega}{2} \sigma_x,
\end{equation}
where $\Delta_0$ is the minimal value of splitting, see Fig.~\ref{fig:fig7}b. Diagonalization of this Hamiltonian yields the energy gap between the qubit eigenstates
\begin{equation}\label{Eq.GAP}
\Delta= \sqrt{\Delta_0^2 + \delta \omega^2}.
\end{equation}
Note that at $\delta\omega \neq 0$ the qubit state precesses about the axis tilted with respect to the main axis of the Bloch sphere, see Fig.~\ref{fig:fig7}d.

\textcolor{NavyBlue}{Expressions \eqref{Eq.bias} and \eqref{Eq.GAP} quantify the relation between the qubit splitting and the electric field paving the way for the realization of  single- and multi-qubit operations with split-ring polariton condensates controlled by the time-dependent external bias. The protocols for their implementation are described in detail in the Appendices~\ref{AppF} and \ref{AppG}. }

\textcolor{NavyBlue}{
\section{Single-qubit operations}\label{AppF}
}

\textcolor{NavyBlue}{
Let us assume that the split-ring qubit is affected by the AC pulse of the radial electric field with the carrier frequency $\omega_d$. It periodically perturbs the persistent currents states whose frequencies are biased according to:
\begin{equation}\label{Eq.driving}
\delta\omega(t) = s(t) \cos(\omega_d t + \phi),
\end{equation}
where $s(t)$ accounts for the pulse envelope and  $\phi$ stands for the relative phase between the driving signal and the qubit.}

\textcolor{NavyBlue}{The effect of the driving can be clearly seen in the frame rotating with the qubit eigenfrequency. To transfer to this frame, we perform the unitary transformation $U_R=\exp\left( -i \Delta \sigma_z t/2\right)$, which transforms the Hamiltonian \eqref{Ham1qb} to $\hat{H}_{\rm rf} = U_R\hat{H} U_R^\dag + i \dot{U}_R U_R^\dag$. After dropping  the fast rotating terms we are left with the following single-qubit Hamiltonian written in the rotating wave approximation (RWA):
\begin{align}\label{HRWA}
\notag \hat{H}_{\rm RWA} = \frac{s(t)}{2}\left(\begin{array}{cc} 0 & e^{i(\omega_d-\Delta_0)t+i \phi}\\
 e^{-i(\omega_d-\Delta_0)t-i \phi} & 0 \end{array}\right) = \\
  \frac{s(t)}{2}\left[ \cos\left((\omega_d-\Delta_0)t+ \phi\right) \sigma_x - \sin\left((\omega_d-\Delta_0)t+ \phi\right) \sigma_y\right].
\end{align}
At the resonance $\omega_d=\Delta_0$, the $\mathcal{R}_x$- and $\mathcal{R}_y$-rotations can be distinguished by the appropriate choice of the relative phase $\phi$. In particular, at $\phi=0$ the qubit rotates about the $x$-axis, while an out-of-phase pulse $\phi=\pi/2$ corresponds to the rotation about the $y$-axis. The rotation angle $\theta$ is determined by the pulse envelope: $\theta = \int{s(t)dt}$.
}

\textcolor{NavyBlue}{In  {Sec.~IIIB} the $Z$-gate was shown to be equivalent to the introduction of time delay between two successive single-qubit operations.  However in practice, especially with many-qubit protocols, this approach fails when one needs to apply a local $\mathcal{R}_z$-gate to the given qubit remaining  the other unaffected. The time interval between the local gates adds a phase to all the qubits in the register, since the time is running equally for all of them. The alternative approach is to shift  the phase of the given qubit with respect to the others by manipulation of the energy gap of the target qubit. At finite $\delta\omega$ the precession axis of the qubit state tilts towards the $x$ axis while the  frequency of precession increases to $\Delta/\hbar$ according to \eqref{Eq.GAP}. The qubit vector precesses faster and accumulates an additional phase shift. Therefore, applying the DC electric pulse on the given qubit, one shifts its relative phase on $\theta(t) = \int_{t^\prime_0}^{t^\prime_0+t} (\sqrt{\Delta_0^2 + \delta\omega(t^\prime)^2} - \Delta_0)dt^\prime$ which is equivalent to the additional $z$-rotation of the qubit state with respect to the unbiased case.}

\textcolor{NavyBlue}{As it was mentioned in the main part of the text, the Pauli rotations can be used for engineering the other gates. The practically important example of a gate synthesis is a Hadamard operation, which performs a $\pi$ rotation about the axis {diagonal in the $x$-$z$ plane}. In particular, it translates the superposition state $\left( |0\rangle + |1\rangle\right)/{\sqrt{2}}$ to the pole of the Bloch sphere. The Hadamard gate can be generated by the sequence of two unitary rotations, namely, the $\pi$ rotation about the $x$ axis followed by the $\pi/2$ rotation about the $y$ axis.
}%
\begin{figure}[h!]
\includegraphics[width=\linewidth]{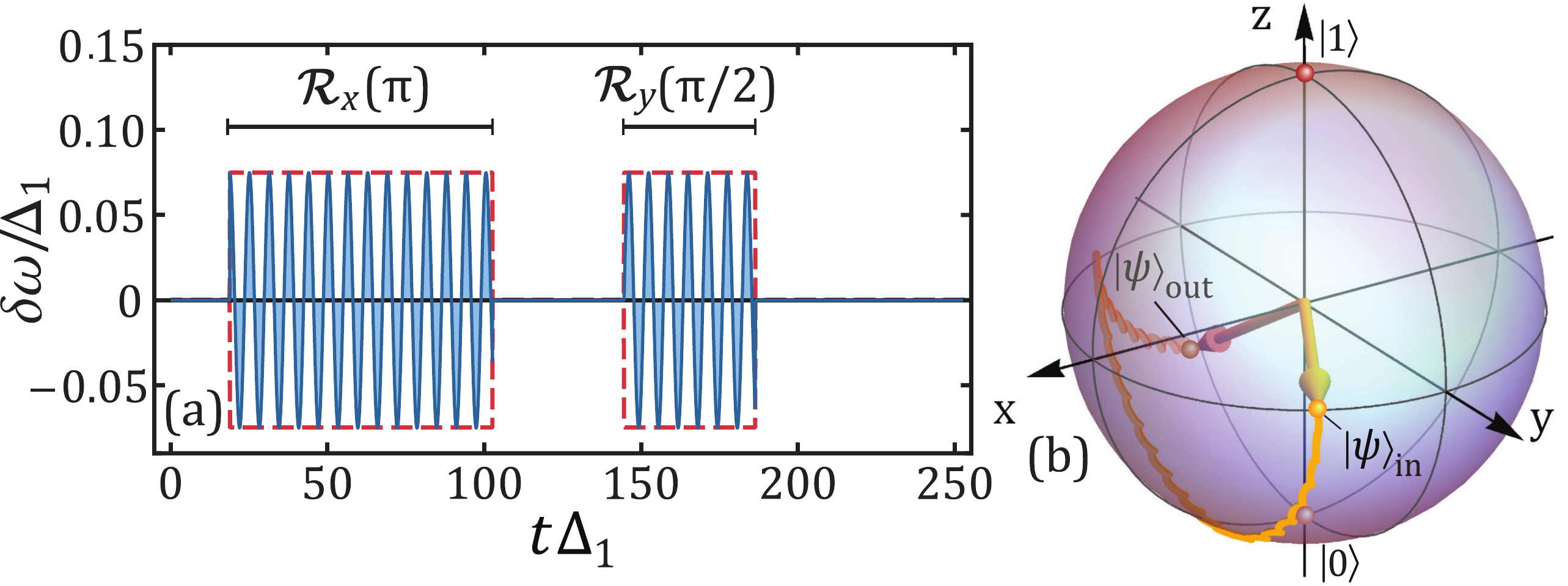}
\caption{Realization of the Hadamard gate composed of two successive Pauli rotations driven by the AC electric pulses (a). The envelope functions are taken to be rectangular. (b) -- Dynamics of the Bloch vector. The initial state taken as $|\psi\rangle_{\rm in} = \left( |0\rangle + e^{i\pi/3} |1\rangle \right)/{\sqrt{2}}$  is shown with the yellow arrow, while the output state $|\psi\rangle_{\rm in}$ corresponds to the red arrow.}\label{Fig.Hadamard}
\end{figure}

\textcolor{NavyBlue}{The scenario of implementation of the Hadamard operation with the approach proposed in this section is shown in Fig.~\ref{Fig.Hadamard}. The shapes of electric pulses responsible for $\mathcal{R}_x$ and $\mathcal{R}_y$ rotations are shown on the left (panel (a)). The Bloch vector dynamics is demonstrated on the panel (b) with the yellow line on the sphere surface. The weak oscillations of the qubit trajectory appear due   to the presence of the fast rotating terms in the total Hamiltonian \eqref{Ham1qb}, which were omitted in the rotating wave approximation~\eqref{HRWA}. The impact of this terms grows as the electric field amplitude increases.
}

\textcolor{NavyBlue}{
\section{The $i$SWAP gate}\label{AppG}
}

\textcolor{NavyBlue}{
The $i$SWAP gate acts to swap the states between the qubits with the addition of $\pi/2$ phase difference. Typically, this gate requires the presence of the $\rm XY$ qubit-qubit interactions \cite{Schuch2003}. Fortunately, this is the kind of coupling, which is realized in the double cavity system shown in Fig.~\ref{fig:fig7}, see Hamiltonian \eqref{2qbHam}. Indeed, in the rotating wave approximation, the evolution operator corresponding to the coupling term reads \cite{Krantz2019}
\begin{eqnarray}\label{Eq.Uint}
\mathcal{U}_{\rm int} &=& e^{-i\frac{g}{2\hbar} \left(\sigma_x\sigma_x + \sigma_y\sigma_y \right)} \notag \\
 &=&
 \left(
\begin{array}{cccc}
 1  & 0 & 0 & 0 \\
0 &  \cos\left(\frac{g}{\hbar}t\right)  & -i\sin\left(\frac{g}{\hbar}t\right) & 0\\
0 & -i\sin\left(\frac{g}{\hbar}t\right) & \cos\left(\frac{g}{\hbar}t\right) & 0\\
0 & 0  & 0 & 1
\end{array}
\right),
\end{eqnarray}
where we assumed that the qubits are in resonance, $\Delta_1=\Delta_2$. As evident from \eqref{Eq.Uint}, at $t=\hbar\pi/2g$ the evolution matrix is identical to the $i$SWAP operation, namely:
\begin{equation}\label{iswap}
\mathcal{U}_{\rm iSWAP} =  \left(
\begin{array}{cccc}
 1  & 0 & 0 & 0 \\
0 &  0  & -i & 0\\
0 & -i & 0 & 0\\
0 & 0  & 0 & 1
\end{array}
\right).
\end{equation}
}

\begin{figure}
\includegraphics[width=1\linewidth]{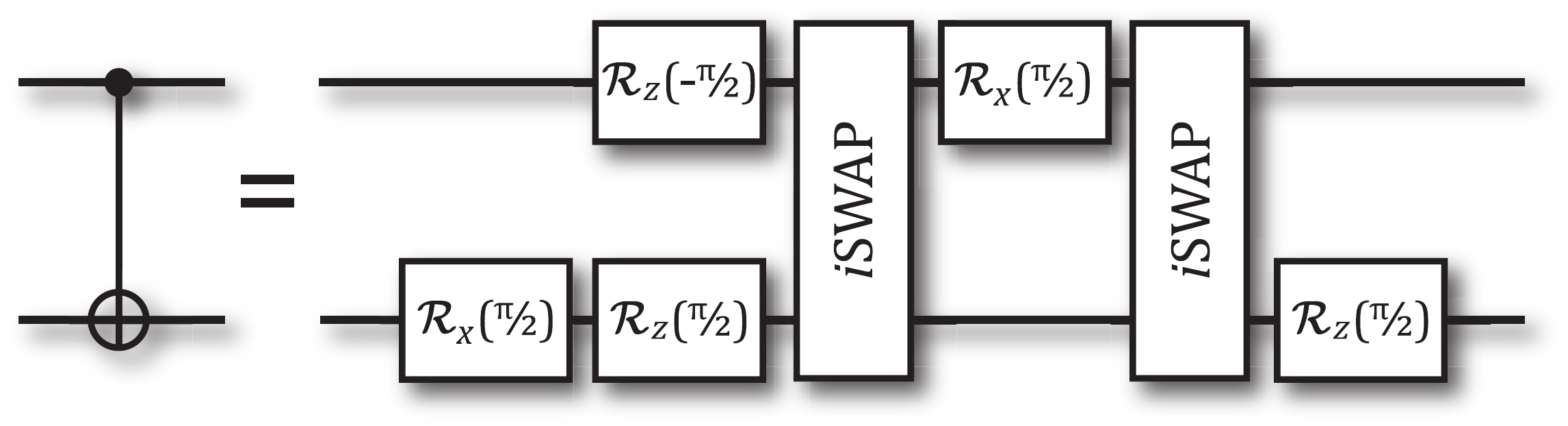}
\caption{Realization of the CNOT-gate with two iSWAPs and several local gates.}\label{Fig.cnot}
\end{figure}

\textcolor{NavyBlue}{
Another important example  of the two-qubit operation is a CNOT-gate, which flips the state of the target qubit conditioned on the control qubit being in state $|1\rangle$, and a zero-controlled NOT-gate (Z-CNOT), which does the same provided that the control qubit is $|0\rangle$. Both these operations can be synthesized upon using  several single- and double qubit gates, see Fig.~\ref{Fig.cnot} and Table~\ref{Table1}. We are particularly interested in these gates as they are required for realization of the Deutsch's algorithm discussed in the following section.
}

\textcolor{NavyBlue}{
\section{Implementaion of the Deutsch's algorithm}\label{AppH}
}

\textcolor{NavyBlue}
{The Deutsch's algorithm is realized with the set of four two-qubit logic functions $\mathcal{U}_i$, whose identities are summarized in Table~\ref{Table1}. The gates corresponding to the constant functions are displayed in the left column. These are the identity gate $\mathcal{U}_1=\mathds{1} \otimes \mathds{1}$ (which performs no operations between the preprocessing and the postprocessing Hadamard transforms of the Deutsch'algorithm, see Fig.~\ref{Fig.DeutschAlgorithm}a) and the second qubit flip gate $\mathcal{U}_2=\mathds{1} \otimes \mathcal{R}_x(\pi)$. The balanced functions $\mathcal{U}_3$ and $\mathcal{U}_4$ are encoded by the CNOT- and the zero-controlled NOT gates, respectively. The circuit shown in Fig.~\ref{Fig.DeutschAlgorithm}a operates with the input state $|\Psi\rangle = (0,1,0,0)^T$, i.e. the first qubit in the state $|0\rangle$ and the second qubit in the state $|1\rangle$.  The algorithm works correctly if the output state of the first qubit is $|0\rangle$ for the gates $\mathcal{U}_1$ and $\mathcal{U}_2$, and $|1\rangle$ for the balanced gates $\mathcal{U}_3$ and $\mathcal{U}_4$.}

\begin{table}
\caption{Definitions of the Oracles for the quantum Deutsch's algorithm. Here we use the following notation for the bit-flip operation $\mathcal{X}=\mathcal{R}_x(\pi)$.\label{Table1}}
\begin{ruledtabular}
\begin{tabular}{ b{0.5cm} b{1.5cm} p{1.5cm}   | b{0.5 cm} b{1.5cm} p{2cm}   }
\multicolumn{3}{c|}{\textbf{Constant functions}} & \multicolumn{3}{c|}{\textbf{Balanced functions}}
\\
\hline
$\mathcal{U}$ & Logic gate  & Circuit  & $\mathcal{U}$ & Logic gate  & Circuit  \\
\hline
$\mathcal{U}_1$
& Identity & {\Qcircuit @C=3em @R=1.5em { & \qw \\ & \qw}} &
 $\mathcal{U}_3$ & CNOT &
 \Qcircuit @C=2.1em @R=1.1em {&  \ctrl{1} & \qw \\ & \targ & \qw}
 \\
\hline
$\mathcal{U}_2$ & Bit-flip & \Qcircuit @C=0.8em @R=1.1em {&  \qw & \qw \\ & \gate{\mathcal{X}} & \qw} &
$\mathcal{U}_4$ & Z-CNOT & \Qcircuit @C=0.3em @R=0.7em {& \gate{\mathcal{X}} &  \ctrl{1} & \gate{\mathcal{X}} \qw & \qw \\ & \qw  & \targ & \qw & \qw} 
\end{tabular}
\end{ruledtabular}
\end{table}

\begin{figure}
\includegraphics[width=\linewidth]{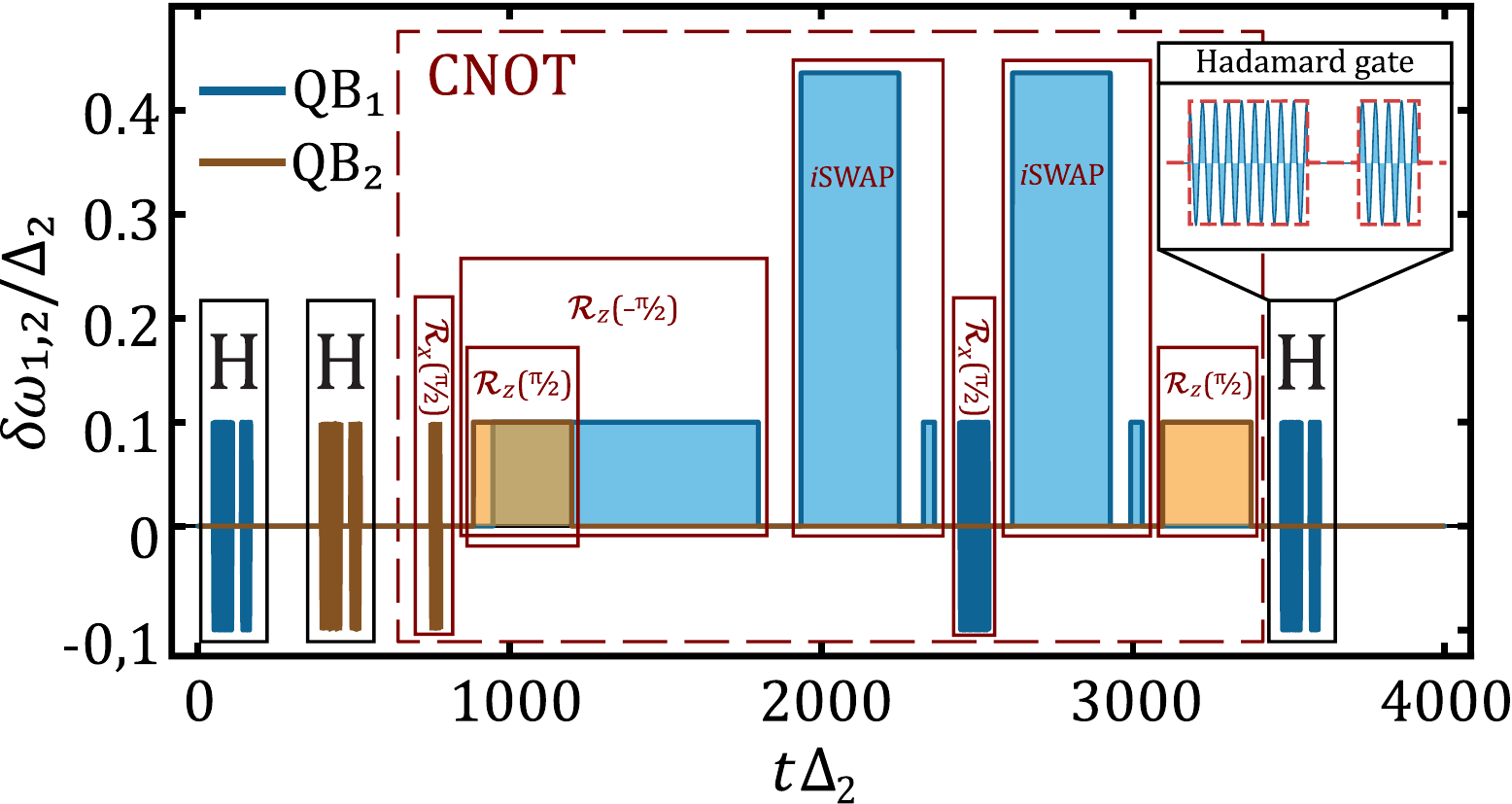}
\caption{The control pulse sequence which encodes a single run of the quantum Deutsch's algortihm with the Oracle $\mathcal{U}_3$. The blue lines show the pulses applied to the first qubit (QB${}_1$) while the orange lines correspond to the pulses governing the second qubit state (QB${}_2$). The Hadamard gates  $\mathbf{H}$  as well as the $\mathcal{R}_x$ operations are implemented by the AC pulses, whose carrier frequency is too high to be resolved at the given temporal scale. For clarification, the signal encoding the last Hadamard operation is zoomed in the inset, see also Fig.~\ref{Fig.Hadamard}. The gap of the second qubit $\Delta_2$ is used as a normalization parameter for both axis grids.}\label{Fig.DA}
\end{figure}

\textcolor{NavyBlue}{
For simulation  of the Deutsch algorithm implementation, we solve the following equation of motion for the 4-component complex two-qubit state column-vector $|\Psi\rangle$, whose elements are the amplitudes of the states $|00\rangle$, $|01\rangle$, $|10\rangle$ and $|11\rangle$:
\begin{equation}
i\partial_t  |\Psi\rangle  = \hat{H}(t)|\Psi\rangle ,
\end{equation}
where $\hat{H}(t)$ is given by Eqs.~\eqref{2qbHam} and \eqref{Eq.GAP}. The sequence of logic gates is encoded in the time dependence of the control parameters $\delta\omega_1(t)$ and $\delta\omega_2(t)$, which are regulated by means of the external voltage, see Fig.~\ref{Fig.DA}. 
We assume that the amplitudes of the AC and DC electric pulses used for the implementation of local Pauli rotations are equal. These values correspond to the bias $\delta\omega_0$, which is considered as the governing parameter for testing the validity of the RWA in Fig.~\ref{Fig.DeutschAlgorithm}b.  The $i$SWAP operations are realized by the square pulses which bring the qubits to the avoided crossing, see Fig.~\ref{fig:fig7}. The amplitude of these pulses is determined by the detuning, $\delta\omega_0=\sqrt{\Delta_2^2 - \Delta_1^2}$. In our simulations we considered the qubits detuned such as $\Delta_1/\Delta_2=0.9$.
}

\textcolor{NavyBlue}{The strongest reduction of the overall algorithm fidelity is due to the limited validity of the rotating wave approximation assumed for the local Pauli gates. Since the gate operation duration is proportional to the electric field amplitude, the use of the weak electric field is undesirable. It makes longer the total run-time of the algorithm, which is fundamentally limited from above by the finite coherence time.}

%

\end{document}